\documentclass[twocolumn]{aastex63}
\hypersetup{linkcolor=red,citecolor=blue,filecolor=cyan,urlcolor=black}

\shorttitle{A first JWST NIRCam photometric catalog}
\shortauthors{Merlin et al.}

\begin{document}

\title{Early results from GLASS-JWST. II. NIRCam Extragalactic Imaging and Photometric Catalog}


\correspondingauthor{Emiliano Merlin}
\email{emiliano.merlin@inaf.it}

\author[0000-0001-6870-8900]{Emiliano Merlin}
\affiliation{INAF Osservatorio Astronomico di Roma, Via Frascati 33, 00078 Monteporzio Catone, Rome, Italy}

\author[0000-0002-2667-5482]{Andrea Bonchi}
\affiliation{Space Science Data Center, Italian Space Agency, via del Politecnico, 00133, Roma, Italy}

\author[0000-0002-7409-8114]{Diego Paris}
\affiliation{INAF Osservatorio Astronomico di Roma, Via Frascati 33, 00078 Monteporzio Catone, Rome, Italy}

\author{Davide Belfiori}
\affiliation{INAF Osservatorio Astronomico di Roma, Via Frascati 33, 00078 Monteporzio Catone, Rome, Italy}

\author[0000-0003-3820-2823]{Adriano Fontana}
\affiliation{INAF Osservatorio Astronomico di Roma, Via Frascati 33, 00078 Monteporzio Catone, Rome, Italy}

\author[0000-0001-9875-8263]{Marco Castellano}
\affiliation{INAF Osservatorio Astronomico di Roma, Via Frascati 33, 00078 Monteporzio Catone, Rome, Italy}

\author[0000-0001-6342-9662]{Mario Nonino}
\affiliation{(INAF - Osservatorio Astronomico di Trieste, Via Tiepolo 11, I-34131 Trieste, Italy)}

\author[0000-0003-4067-9196]{Gianluca Polenta}
\affiliation{Space Science Data Center, Italian Space Agency, via del Politecnico, 00133, Roma, Italy}

\author[0000-0002-9334-8705]{Paola Santini}
\affiliation{INAF Osservatorio Astronomico di Roma, Via Frascati 33, 00078 Monteporzio Catone, Rome, Italy}

\author[0000-0002-8434-880X]{Lilan Yang}
\affiliation{Kavli Institute for the Physics and Mathematics of the Universe, The University of Tokyo, Kashiwa, Japan 277-8583}

\author[0000-0002-3254-9044]{Karl Glazebrook}
\affiliation{Centre for Astrophysics and Supercomputing, Swinburne
University of technology, P.O. Box 218, Hawthorn, VIC 3122, Australia}

\author[0000-0002-8460-0390]{Tommaso Treu}
\affiliation{Department of Physics and Astronomy, University of California, Los Angeles, 430 Portola Plaza, Los Angeles, CA 90095, USA}

\author[0000-0002-4140-1367]{Guido Roberts-Borsani}
\affiliation{Department of Physics and Astronomy, University of California, Los Angeles, 430 Portola Plaza, Los Angeles, CA 90095, USA}

\author[0000-0001-9391-305X]{Michele Trenti}
\affiliation{School of Physics, University of Melbourne, Parkville 3010, VIC, Australia}
\affiliation{ARC Centre of Excellence for All Sky Astrophysics in 3 Dimensions (ASTRO 3D), Australia}

\author[0000-0003-3195-5507]{Simon Birrer}
\affiliation{Kavli Institute for Particle Astrophysics and Cosmology and Department of Physics, Stanford University, Stanford, CA 94305, USA}
\affiliation{SLAC National Accelerator Laboratory, Menlo Park, CA, 94025}
\affiliation{Department of Physics and Astronomy, Stony Brook University, Stony Brook, NY 11794, USA}

\author[0000-0003-2680-005X]{Gabriel Brammer}
\affiliation{Cosmic Dawn Center (DAWN), Denmark}
\affiliation{Niels Bohr Institute, University of Copenhagen, Jagtvej 128, DK-2200 Copenhagen N, Denmark}

\author[0000-0002-5926-7143]{Claudio Grillo}
\affiliation{Dipartimento di Fisica, Università degli Studi di Milano, via Celoria 16, I-20133 Milano, Italy}
\affiliation{INAF—IASF Milano, via A. Corti 12, I-20133 Milano, Italy}

\author[0000-0003-2536-1614]{Antonello Calabr\`o}
\affiliation{INAF Osservatorio Astronomico di Roma, Via Frascati 33, 00078 Monteporzio Catone, Rome, Italy}

\author[0000-0001-9002-3502]{Danilo Marchesini}
\affiliation{Department of Physics and Astronomy, Tufts University, 574 Boston Ave., Medford, MA 02155, USA}

\author[0000-0002-3407-1785]{Charlotte Mason}
\affiliation{Cosmic Dawn Center (DAWN), Denmark}
\affiliation{Niels Bohr Institute, University of Copenhagen, Jagtvej 128, 2200 København N, Denmark}

\author{Amata Mercurio}
\affiliation{INAF -- Osservatorio Astronomico di Capodimonte, Via Moiariello 16, I-80131 Napoli, Italy}

\author[0000-0002-8512-1404]{Takahiro Morishita}
\affiliation{IPAC, California Institute of Technology, MC 314-6, 1200 E. California Boulevard, Pasadena, CA 91125, USA}

\author[0000-0002-6338-7295]{Victoria Strait}
\affiliation{Cosmic Dawn Center (DAWN)}
\affiliation{Niels Bohr Institute, University of Copenhagen, Jagtvej 128, 2200 København N, Denmark}

\author[0000-0003-4109-304X]{Kristan Boyett}
\affiliation{School of Physics, University of Melbourne, Parkville 3010, VIC, Australia}
\affiliation{ARC Centre of Excellence for All Sky Astrophysics in 3 Dimensions (ASTRO 3D), Australia}

\author[0000-0003-4570-3159]{Nicha Leethochawalit}
\affiliation{School of Physics, University of Melbourne, Parkville 3010, VIC, Australia}
\affiliation{ARC Centre of Excellence for All Sky Astrophysics in 3 Dimensions (ASTRO 3D), Australia}
\affiliation{National Astronomical Research Institute of Thailand (NARIT), Mae Rim, Chiang Mai, 50180, Thailand}

\author[0000-0003-2804-0648 ]{Themiya Nanayakkara}
\affiliation{Centre for Astrophysics and Supercomputing, Swinburne University of Technology, PO Box 218, Hawthorn, VIC 3122, Australia}

\author[0000-0003-0980-1499]{Benedetta Vulcani}
\affiliation{INAF Osservatorio Astronomico di Padova, vicolo dell'Osservatorio 5, 35122 Padova, Italy}


\author[0000-0001-5984-0395]{Marusa Bradac}
\affiliation{University of Ljubljana, Department of Mathematics and Physics, Jadranska ulica 19, SI-1000 Ljubljana, Slovenia}
\affiliation{Department of Physics and Astronomy, University of California Davis, 1 Shields Avenue, Davis, CA 95616, USA}

\author[0000-0002-9373-3865]{Xin Wang}
\affil{Infrared Processing and Analysis Center, Caltech, 1200 E. California Blvd., Pasadena, CA 91125, USA}

\begin{abstract}
We present the reduced images and multi--wavelength catalog of the first JWST NIRCam extragalactic observations from the GLASS Early Release Science Program, obtained as coordinated parallels of the NIRISS observations of the Abell 2744 cluster. Images in seven bands (F090W, F115W, F150W, F200W, F277W, F356W, F444W) have been reduced using an augmented version of the official JWST pipeline; we discuss the procedures adopted to remove or mitigate defects in the raw images. We obtain a multi--band catalog by means of forced aperture photometry on PSF-matched images at the position of F444W-detected sources. The catalog is intended to enable early scientific investigations, and it is optimized for faint galaxies; it contains 6368 sources, with limiting magnitude 29.7 at 5$\sigma$ in F444W. We release both images and catalog in order to allow the community to familiarize with the JWST NIRCam data and evaluate their merit and limitations given the current level of knowledge of the instrument. 
\end{abstract}


\section{Introduction}
\label{sec:intro}

The first James Webb Space Telescope scientific images have been made available to the public on July 14th, 2022. Among them, the GLASS Early Release Science (GLASS-ERS) Program \citep[JWST-ERS-1324, PI Treu; see][T22 hereafter]{Treu2022} has obtained the deepest ERS spectro-photometric data, observing the region of the Hubble Frontier Field galaxy cluster Abell 2744 with three instruments. The first GLASS NIRCam dataset \citep[see][for a presentation of the instrument]{Burriesci2005, Rieke2005} was obtained on June 28-29th, 2022, and it consists of images in seven bands from a parallel pointing of the NIRISS field of view (FoV) on the cluster, with an area of $\sim$9 sq. arcmin. The FoV is centered at RA=3.5017025 deg, Dec=-30.3375436 deg, i.e. at around one virial radius ($\sim$1--2.5 Mpc) from the cluster center. We refer the reader to Section 6.1 of T22 for full details on the observational setup.


This work is part of a series of short letters illustrating the methods and results from this very first JWST data release. Paper I \citep{RobertsBorsani2022} focuses on NIRISS data, while
with this paper we provide the community with an initial release of these NIRCam processed data to allow for early investigation in accompanying first science papers and preparation of future work with NIRCam. 
This work will also form the basis for a larger and more comprehensive data release when additional NIRCAM imaging is taken for JWST-ERS-1324.

The processing of NIRCam images has been performed with a customised version of the STScI pipeline, adopting the reference and calibration data available at the moment of writing this paper\footnote{The images were initially processed using the files available at the moment of their acquisition, and were re-processed during the revision of the paper using the new files made available on 2022 July 29: \texttt{CAL\_VER 1.6.0}; \texttt{CRDS\_CTX jwst\_0942.pmap}.}, and is therefore preliminary. This work provides the first users of NIRCam public data with an evaluation of the steps required to reduce the data, the level of uncertainties in their calibration and initial mitigation strategies for the removal of instrumental defects. 

We then used well-tested techniques, inherited from previous projects on deep surveys such as CANDELS \citep{Grogin2011, Koekemoer2011,Galametz2013} and ASTRODEEP \citep{Merlin2016b,Castellano2016,Merlin2021}, to produce a first photometric catalog, and we are releasing it along with the mosaics of the observed FoV. 

We note that modest lensing magnification is expected to be present in the FoV. 
In this initial set of papers we neglect the effect; the issue will be revisited after the completion of the campaign. However, this has no direct impact on the present work, as we are only releasing measured fluxes. 

Despite the mentioned uncertainties, these results provide an illustration of the power of JWST in exploring the distant Universe. We expect the quality of the processing to improve over the next months, thanks to the progress of calibration activities and the refinement of analysis techniques.

This paper is organized as follows. In Section \ref{sec:data} we present the data-set and discuss the image processing pipeline. In Section \ref{sec:detection} the methods applied for the detection of the sources are described, and in Section \ref{sec:photometry} we present the photometric techniques used to compute the fluxes. Section \ref{sec:caveats} lists some known caveats on the accuracy of the catalog, and in Section \ref{sec:conclusions} we summarize the results. Throughout the paper we adopt AB magnitudes \citep{oke83}.


\section{Data reduction}
\label{sec:data}

The present dataset consists of seven NIRCam bands images, from 0.9 to 4.4$\mu$m. The FoV is divided in two separated modules. Four images are ``short-wavelength'': F090W, F115W, F150W, and F200W, with typical native pixels scale close to 0.031$\arcsec$, and each module observed with four detectors. The other three are ``long wavelength'': F277W, F356W, and F444W, with typical native pixel scale close to 0.063$\arcsec$, and a single detector per module. We used a customised version of the official STScI JWST pipeline\footnote{\url{https://jwst-docs.stsci.edu/jwst-science-calibration-pipeline-overview}} to create mosaics starting from the raw images, with changes motivated by the analysis of tests we performed on simulated data.

We produced the simulations using \textsc{Mirage}\footnote{\url{https://www.stsci.edu/jwst/science-planning/proposal-planning-toolbox/mirage}}, a software package which creates synthetic NIRCam raw images with celestial position and depth consistent with the actual scheduling of the Program as derived from the JWST Astronomer's Proposal Tool. To populate the simulated images, \textsc{Mirage} requires an input catalog which we produced using \textsc{Egg} \citep[][]{schreiber2017}, a software that uses empirical relations calibrated on the CANDELS data to create mock galaxy catalogs. The sources are simulated as two-component objects, with a bulge and a disk both described by \citet{Sersic1968} profiles with $n=4$ and 1, respectively. We also included an additional population of high-redshift galaxies to allow for the study of color selection criteria \citep[see Paper III,][]{Castellano2022},
a stellar field obtained using the \textsc{Trilegal} application \citep{Girardi2005,Girardi2012}, and bright stars at the positions of the GAIA sources falling in the FoV to compute the astrometric solution. We used these simulations to prepare the reduction process, introducing some of the improvements described below,  and validate the photometric measurement methods.

\begin{figure}
\center
 \includegraphics[width=0.46\textwidth]{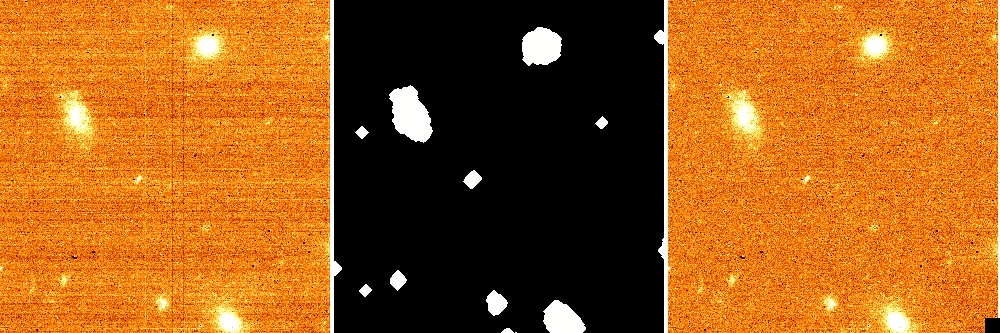}
 
 \vspace{0.25cm}
 
 \includegraphics[width=0.46\textwidth]{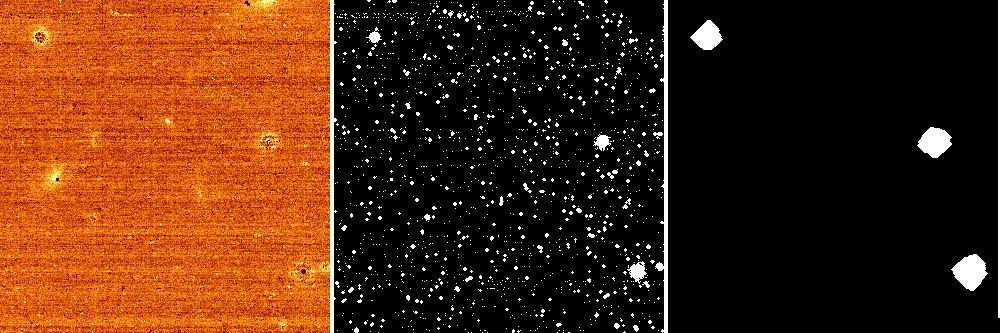}
 
 \vspace{0.25cm} 

\includegraphics[width=0.15\textwidth]{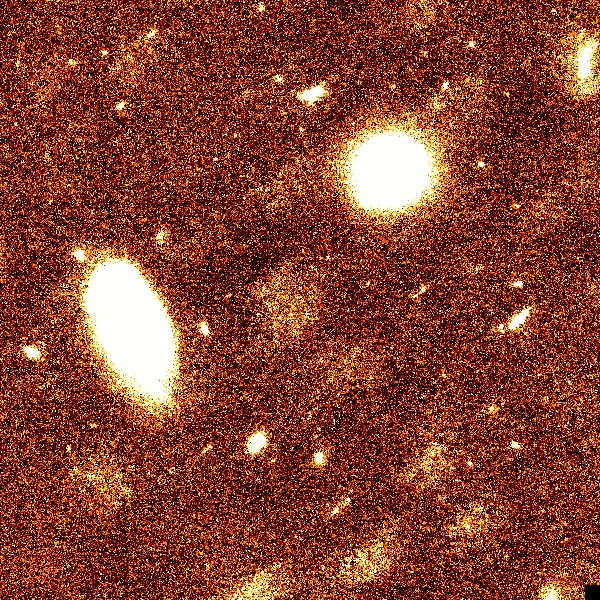}
\includegraphics[width=0.15\textwidth]{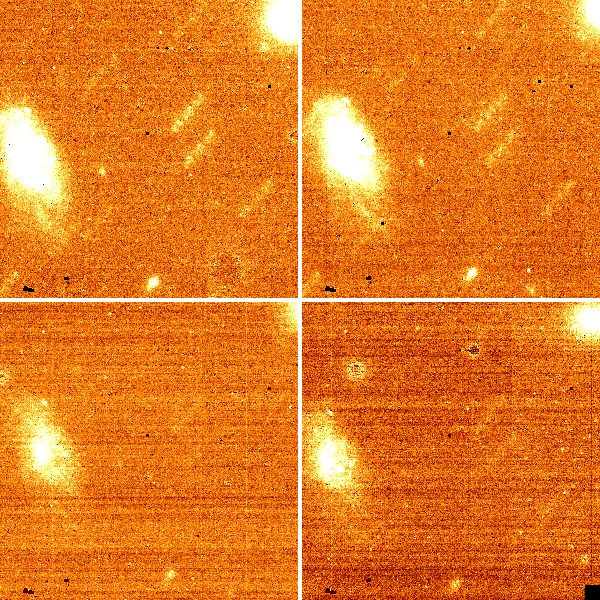}
\includegraphics[width=0.15\textwidth]{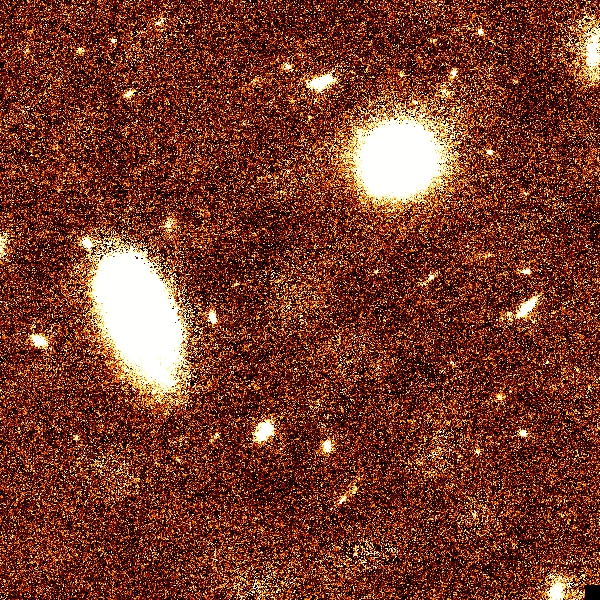} 
 
 \caption{Correction of defects in the raw images. \textbf{Top panels}: Effect of the subtraction to remove the 1/f noise in both vertical and horizontal directions. \textit{Left}: original calibrated image; \textit{Center}: dilated mask to remove objects and bad pixels from the estimate of the median row/column level; \textit{Right}: Image after noise removal. \textbf{Middle panels}: Effect of the method adopted to remove the ``snowballs'' from the images. \textit{Left}: original image; \textit{Center}: data quality image as produced by the first two stages of the STScI pipeline. Snowballs are identified as large clusters of pixels typically flagged with value 4. \textit{Right} Mask obtained by the procedure described in the text. This example is drawn from one image in the F115W filter. \textbf{Bottom panels}: an illustration of the presence of scattered light due to a ``wing-tilt'' event during observation which caused the offset / doubling of sources in three detectors of the F090W and F115W data, and results of the  tentative masking and removal procedure. This example deliberately shows the worst-case scenario, i.e. a portion of the b4 chip in the F090W filter. The effect is much milder or absent in other modules and at wavelengths longer than F150W. \textit{Left}: Mosaic of all F090W images without any masking/removal. \textit{Center}: Individual frames with scattered light patterns. They are more evident in the two longer exposure (upper panels). \textit{Right}: Mosaic of all F090W images after masking of pixels affected by scattered light following the method described in the text.}
 \label{imgproc}    
\end{figure}

Real data were then processed as follows. From raw exposures, we used Levels 1 and 2 of the STScI pipeline (see T22 Appendix) to apply flat--field and dark current correction, and obtain data quality masks (DQM) to identify and potentially flag other detector-level defects. 
We then applied a number of custom procedures to remove three types of instrumental defects that are not dealt with by the currently available STScI pipeline. Specifically, we have applied corrections for:

\begin{itemize}
\item \textit{1/f noise}, which introduces random vertical and horizontal stripes into the images \citep[see][]{Schlawin2020}. It has been removed by subtracting the median value from each line/column, after masking out all the objects and the bad pixels. The masks have been obtained by running \textsc{SExtractor} \citep{Bertin1996} and then dilating the resulting segmentation image by 15 pixels. See the upper panels of Fig. \ref{imgproc};

\item \textit{``Snowballs''}, i.e. circular defects that have been observed in after-launch data \citep[see][]{Rigby2022}. These are identified in the DQM, but they are not flagged as ``do not use'' pixels. Because these defects appear as large clusters of pixels in the DQM, we were able to detect them using again \textsc{SExtractor} on the DQM itself, setting suitable large threshold areas and retaining only objects with low ellipticity. We have found that an area of 80 pixels is typically effective in identifying 90\% of the snowballs in the DQM; we note that this mask is somewhat conservative, as it sometimes identifies regions of unaffected pixels. The masked pixels are excluded from the final sum when creating the scientific mosaic. The procedure is illustrated in the middle panels of Fig.~\ref{imgproc};

\item \textit{Scattered light}: as shown in  the lower panels of Fig.~\ref{imgproc}, we have found several additive features in F090W and in F115W (in three detectors only, namely a3, b3 and b4). After investigation by JWST technicians during the revision of this paper, it was found out that the issue is due to a so-called ``wing-tilt event'' that happened during the observations; in practice, a small shift in position of one of the wings of the primary mirror caused the offset/doubling of sources in the data. A full re-observation of the affected bands was approved by the STScI, and the new images will be used for future versions of the catalog. For the present release, we have applied the following procedure to the F090W and F115W frames to identify and remove the spurious structures. We run \textsc{SExtractor} on a median average of the single exposures to obtain a segmentation map containing both real sources and defects, and then on the F444W image where the defects are not present. Cross-correlating the two maps, we were able to single out most of the spurious objects and to mask out their pixels from the final sum, without removing real objects. In the cases where the effect was particularly strong, the final result is a clear improvement of the image quality, even if some residual scattered light is present; see the lower panels of Fig.~\ref{imgproc}.

\end{itemize}

Clearly, these procedures should be considered as preliminary and temporary means to alleviate these problems, pending definitive solutions that will be implemented in future releases of the STScI pipeline. 

The astrometric calibration was performed using \textsc{SCAMP} \citep[][]{Bertin2006}, with 3rd order distortion corrections (\texttt{PV} coefficients up to $j=10$), in two steps: first, we aligned the F444W band single exposures to ground-based catalogs (obtained in the $i$ band with the Magellan telescope in good seeing condition, see T22) of the same region, which had been previously aligned to GAIA-DR3 stars \citep[][2022 in prep.]{Prusti2016}; then, we took the resulting high-resolution catalog in F444W as reference for the other JWST bands, using compact, isolated sources detected at high signal-to-noise at all wavelengths. Each NIRCam detector has been analysed independently, in order to simplify the treatment of distortions and minimise the offsets of the sources in different exposures. The average difference between the positions of bright sources retrieved with \textsc{SExtractor} runs on each band with respect to the positions in the master catalog used in the \textsc{SCAMP} runs is of $\sim$1 mas, with r.m.s. scatter of 15 mas. In some small regions of the short wavelength images (particularly close to the borders of the detectors) the alignment was sub-optimal even after this procedure, so we masked them out from the final mosaics.

We then rescaled the single exposures to units of $\mu$Jy/pixel, using the conversion factors outputted by the pipeline. 
Once again we warn the reader that we used the currently available preliminary calibrations. 
We checked that the zero-points are reasonable, comparing the measured fluxes of bright sources in four bands (F090W, F150W, F356W and F444W) with archival ground-based and \textit{Spitzer} data; 
finding overall consistency within 0.1-0.2 magnitudes (see Section \ref{sec:photometry}). It is difficult to obtain more precise indications, given the difference in resolution and depth between the instruments. Similar results were obtained by checking the conversion factors given by the pipeline with the reference numbers listed in \citet{Rigby2022}, Table 3. Of course, the zero-points will be calibrated more accurately when updated configuration files are available. 

Finally, we used \textsc{SWarp} \citep[][]{Bertin2002} to combine the single exposures into mosaics projected onto a common aligned grid of pixels, and \textsc{SExtractor} to further clean the images by subtracting the residual sky background. The pixel scale of all the images was set to 0.031$\arcsec$ (the approximate native value of the short wavelength bands), to allow for simple processing with photometric algorithms. 

We made a first estimate of the depths of the mosaics by injecting artificial point sources (i.e. PSF stamps, see below) of known magnitude in empty regions, and measuring their flux and uncertainties with \textsc{a-phot} \citep{Merlin2019}, using apertures of radius 0.1\arcsec as in Table 2 from T22; we found good agreement with their values. We then fine-tuned the RMS maps produced by the pipeline, rescaling them by appropriate multiplicative factors (with values ranging from 1.08 to 1.34), to make them fully consistent with the dispersion of the measured fluxes of the artificial sources, which provides an accurate estimate of the real uncertainties including correlated errors between pixels. After this correction, the maximum depths of most of the bands end up being slightly shallower (up to $\sim$0.3 magnitude) than the ones in T22, while F444W is slightly deeper. In Table \ref{parameters} we list the final reference depths, estimated as the mode of the limiting total magnitudes (5$\sigma$ in 0.1\arcsec radius apertures) corresponding to the value of each pixel in the corrected RMS maps, along with the total exposure times and the FWHM of the PSF for each band.

\begin{figure*}
\center
 \includegraphics[width=\textwidth]{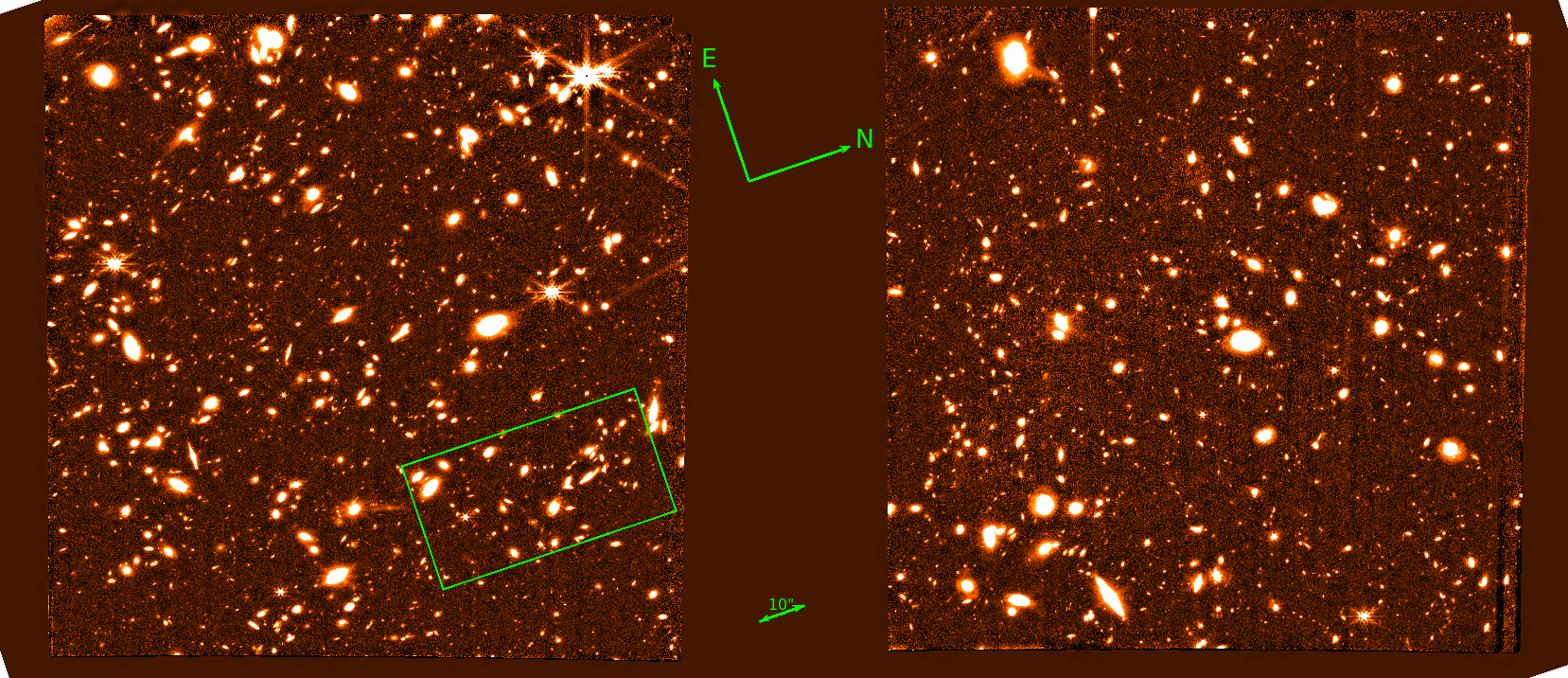}
 \includegraphics[width=\textwidth]{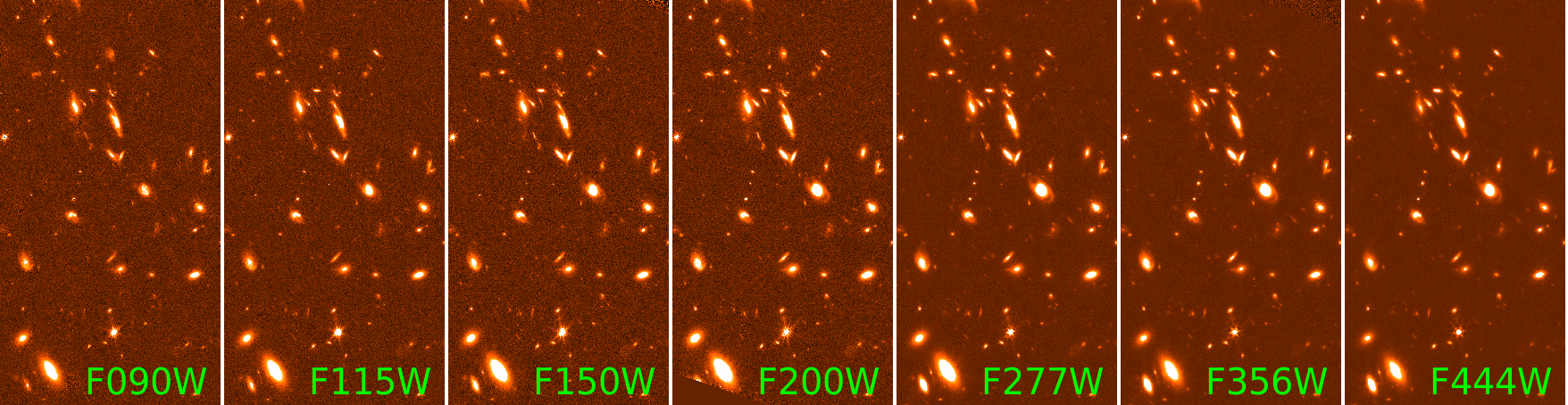}
  \includegraphics[width=\textwidth]{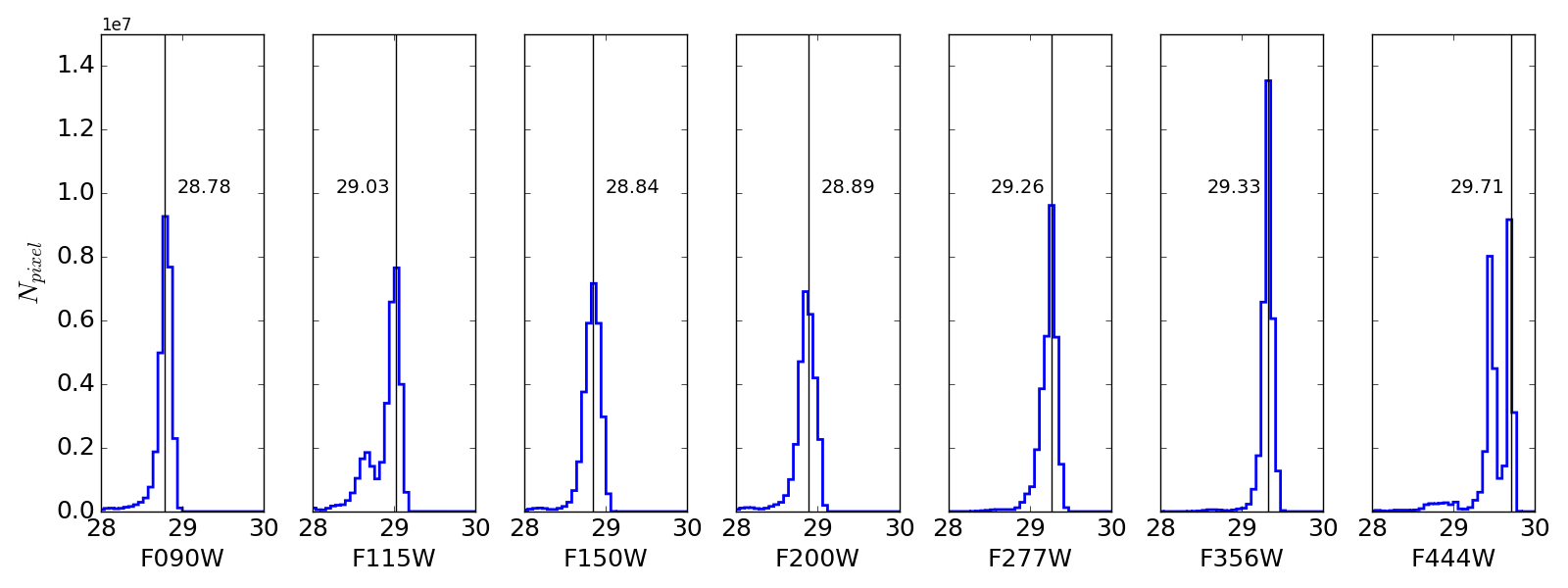}
 \caption{Top panel: full view of the F444W mosaics. Module \textit{b} is shown on the left, module \textit{a} on the right.  Middle panels:  the small region in the green box of the FoV ($25\arcsec \times 48 \arcsec$) in the seven observed bands (left to right, F090W, F115W, F150W, F200W, F277W, F356W, and F444W); the color cut is the same in all bands. A three-color composite image of the field is available online (see Sect. \ref{sec:conclusions}). Bottom panels: histograms of the 5$\sigma$ limiting magnitude for all pixels in each mosaic, computed using the rescaled RMS maps as described in Sect. \ref{sec:data}. Note the double peak in F444W, caused by the slightly different depth of the two modules.}
 \label{fig1}
\end{figure*}

Figure \ref{fig1} shows the final F444W full mosaic, and a small region of the FoV in the seven bands. Clearly, the processing is still not perfect; most noticeably, it shows some residual background features in module  \textit{a} (rightmost in the image). We expect that these features will be further reduced as reference files and pipeline are upgraded. However, we find it good enough to perform a first round of scientific analysis, keeping in mind its limitations. The bottom panels of Fig. \ref{fig1} show the distributions of the values of limiting total magnitude of all pixels in each band, computed as described above, and the mode of the distribution. 

\begin{table}
    \centering
    \begin{tabular}{cccc}
        \hline
        Band & FWHM & Exposure time & Depth \\
        \hline
        \hline 
        F090W & 0.035 & 11520 & 28.78 \\
        \hline
        F115W & 0.040 & 11520 & 29.03 \\
        \hline
        F150W & 0.050 & 6120 & 28.84 \\
        \hline
        F200W & 0.065 & 5400 & 28.89 \\
        \hline
        F277W & 0.095 & 5400 & 29.26 \\
        \hline
        F356W & 0.115 & 6120 & 29.33 \\
        \hline
        F444W & 0.140 & 23400 & 29.71 \\
        \hline
    \end{tabular}
    \caption{Resolution and maximum depth of the seven bands in the data-set. FWHM are in arcsec; exposure times in seconds; depths are 5$\sigma$ point source magnitudes in 0.1$\arcsec$ radius apertures.}
    \label{parameters}
\end{table}

\section{Detection}
\label{sec:detection}

We performed detection on the F444W band. The reasons for this choice is twofold: F444W is the deepest among the seven bands, and 
high-redshift sources (which are the main targets for this first round of studies) typically have the brightest flux in the reddest band. In contrast, F444W has the broadest FWHM (0.14$\arcsec$) of the set, so in principle - given the high density of objects - contamination could be an issue. However, visually inspecting the segmentation map we found that faint objects are typically well-isolated, and confusion is not too  significant. 

\begin{table}
    \centering
    \begin{tabular}{lc}
        \hline
        Parameter & value\\
        \hline
        \hline 
        \texttt{DETECT\_MINAREA} & 8 \\
        \hline
        \texttt{DETECT\_THRESH} & 0.7071\\
        \hline
        \texttt{ANALYSIS\_THRESH} & 0.7071\\
        \hline
        \texttt{DEBLEND\_NTHRESH} & 32\\
        \hline
        \texttt{DEBLEND\_MINCOUNT} & 0.0003\\
        \hline
        \texttt{BACK\_SIZE} & 64\\
        \hline
        \texttt{BACK\_FILTERSIZE} & 3\\
        \hline
    \end{tabular}
    \caption{\textsc{SExtractor} parameters used for the detection procedure on the F444W mosaic.}
    \label{SExtab}
\end{table}

\begin{figure}
\center
\includegraphics[width=0.42\textwidth]{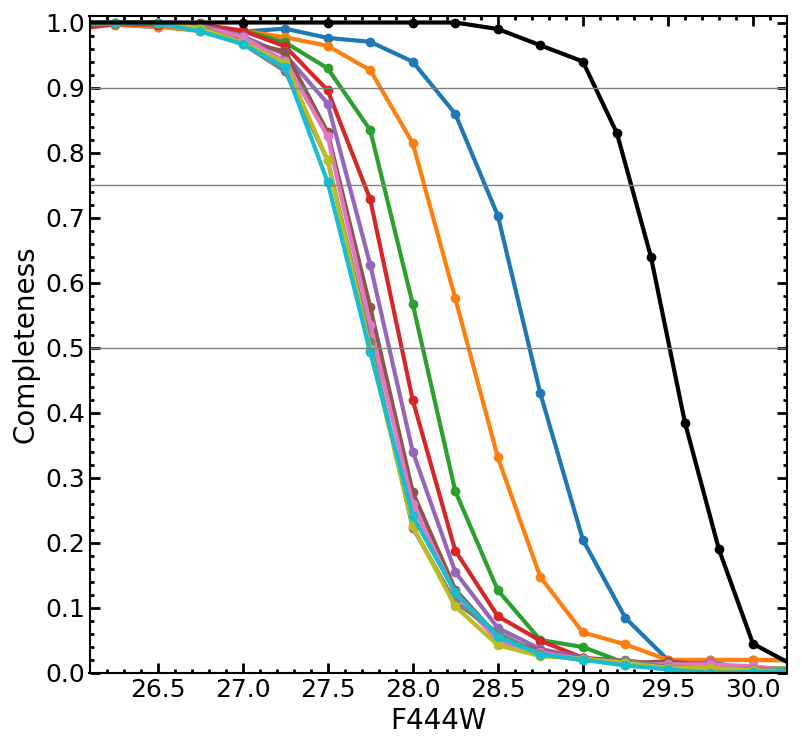}
\includegraphics[width=0.47\textwidth]{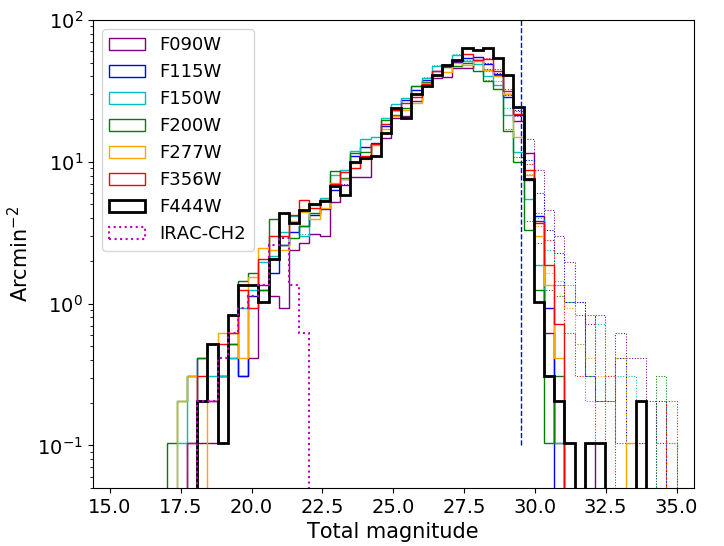}
\caption{Top panel: detection completeness as a function of the F444W magnitute, estimated by injecting artificial sources in empty regions of the mosaic and detecting with \textsc{SExtractor}. The colored lines correspond to disks with half-light radii going from 1$\arcsec$ (leftmost cyan line) to 0.1$\arcsec$ (rightmost blue line); the black line is for point sources. Horizontal thin lines mark the 50\%, 75\% and 90\% completeness level.
Bottom panel: number counts (sources per sq. arcmin) of the detected sources, as a function of their measured total magnitude (computed using colors in 2 FWHMs, see Sect. \ref{sec:photometry}), in all bands. We included all the detections, without attempting at isolating stars or spurious detections, which anyway are a small fraction of the total. The dashed vertical blue line marks the 50\% completeness limit in detection. For comparison, we also show the \textsc{Spitzer} IRAC 4.5 $\mu$m counts on the same region, obtained by running \textsc{SExtractor} on the Hubble Frontier Fields images by \citet{Lotz2017}.} 
 \label{counts}
\end{figure}

We used \textsc{SExtractor} v2.8.6 in the customised version used for the CANDELS campaign \citep[see][]{Galametz2013}, smoothing the scientific image with a Gaussian convolution filter with FWHM=0.14$\arcsec$, and applying a detection threshold corresponding to a signal-to-noise ratio (SNR) of 2. This quite aggressive choice was made on the basis of the simulations, to obtain a good balance between completeness and purity. We checked by visual inspection that it allows for the detection of faint sources that are consistently detected also in F356W, without apparently including many spurious ones. We estimated the detection completeness by injecting artificial sources of known magnitude and different morphology (point-sources and face-on disks with half-light radii 0.1$\leq R_h \leq$1.0 arcsec) in empty regions of the mosaic (to factorize out other selection effects due to e.g. the choice of detection and deblending parameters), and then running \textsc{SExtractor} again with the same parameters to check which fraction of them are detected. 
With this technique we find a completeness of 90\% at F444W=29.1, of 75\% at F444W=29.3, and of 50\% at F444W=29.5 for point-sources. The upper panel of Fig. \ref{counts} shows the completeness for the different classes of injected sources as a function of the input magnitude. On the simulated data, with comparable background levels and the same PSF models, we obtained a purity of $\sim$97\% all detections down to the 50\% completeness magnitude; we warn that this is an optimistic estimate, considering the many uncertainties and anomalies found we found in the real images that were not present in the simulations. 

The other parameters used in the run are listed in Table \ref{SExtab} (we include the values for background which are relevant to the detection process, but we stress that the background subtraction has been performed during image processing as described in Sect. \ref{sec:data}).
The final \textsc{SExtractor} catalog contains 6368 objects. For this first release we did not attempt any cleaning of spurious detections or wrong deblendings.

\section{Photometry}
\label{sec:photometry}

To measure the fluxes and colors of the detected objects in all bands, we followed a strategy similar to that adopted for Hubble Space Telescope (HST) images in CANDELS \citep[see e.g.][]{Galametz2013} and in ASTRODEEP \citep{Merlin2016b, Merlin2021}. Since the detection band is the one with the coarsest resolution, we PSF-matched all the other images to it for color fidelity. We created convolution kernels using 
the \textsc{WebbPSF} models publicly provided by STScI\footnote{\url{https://jwst-docs.stsci.edu/jwst-near-infrared-camera/nircam-predicted-performance/nircam-point-spread-functions}}, combining them with a Wiener filtering algorithm based on the one described in \citet{Boucaud2016}; and we used a customised version of the convolution module in \textsc{t-phot} \citep{Merlin2015,Merlin2016a}, which uses \texttt{FFTW3} libraries, to smooth the images. We note that we chose to use these PSFs after trying to create models from the few unsaturated stars available in the fields, using the software \textsc{Galight}.  \citep{Ding2020}; we checked that the FWHMs of the resulting PSFs were consistent with the those of the \textsc{WebbPSF} ones, but the resulting convolution kernels were too noisy and yielded more scattered color estimations. The resulting PSF-matched images look good at visual inspection, with no evident signs of artefacts introduced by the procedure, even in the case of the bands with FWHM close to that of F444W.
Then, we used \textsc{a-phot} to measure the fluxes at the positions of the detected sources on the PSF-matched images, masking neighboring objects using the \textsc{SExtractor} segmentation map. We measured 
the flux within the segmentation area (the images being on the same grid and PSF-matched), and the fluxes within six circular apertures with 0.2$\arcsec$, 0.28$\arcsec$, 0.42$\arcsec$, 0.56$\arcsec$, 1.12$\arcsec$, and 2.24$\arcsec$ diameters (corresponding to 2 to 16 FWHMs). 

\begin{figure}
\center
\includegraphics[width=0.47\textwidth]{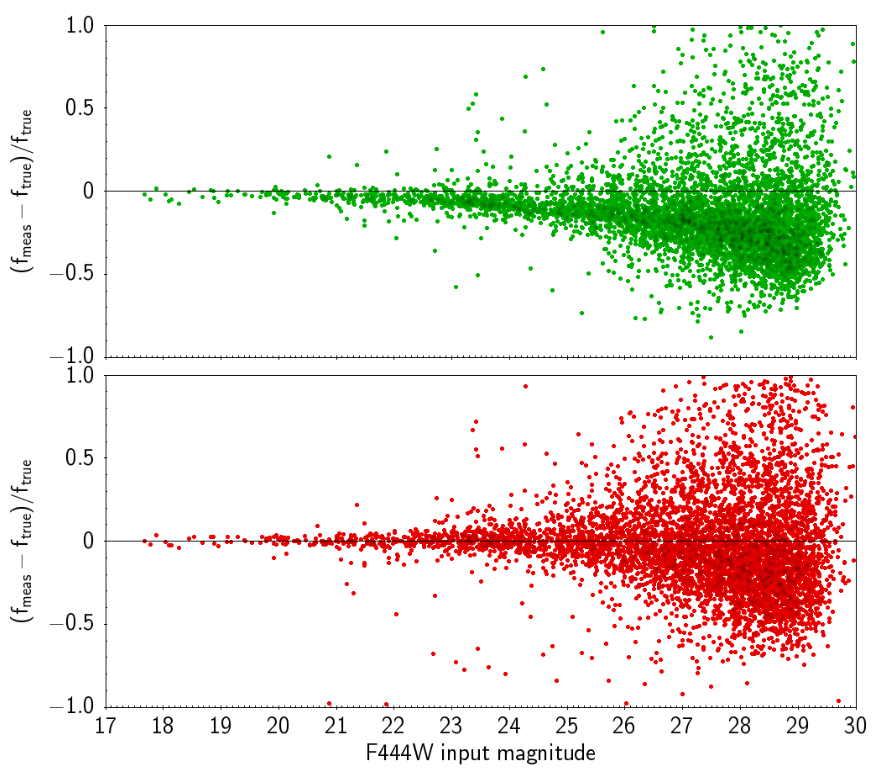}
\caption{
Comparison between \textsc{SExtractor} \texttt{FLUX\_AUTO} (green) and \textsc{a-phot} (red) estimated Kron flux on a simulated F444W image; the plots show the relative error with respect to the input (simulated) flux as a function of the input magnitude, and demonstrate that the \textsc{a-phot} estimate yields an overall lower median bias.} 
 \label{sim}
\end{figure}

On the detection image F444W we also estimated a total flux by means of a Kron elliptical aperture \citep{Kron1980}. We point out that while the \textsc{a-phot} Kron flux is conceptually identical to the \textsc{SExtractor} one, we use different parameters which on the simulated data yielded a smaller median bias at the expense of a slightly larger dispersion. This is shown in 
Fig. \ref{sim}, where the relative error between measured and input fluxes in a simulated F444W field are shown for both cases. 
We also want to stress that while the Kron estimate is a good proxy for the total flux of an object, it is prone to errors due to contamination from nearby sources, even after masking them out using the detection segmentation map (because the segmented area typically does not include faint extended wings). Detailed studies focused on individual objects should therefore take this estimate of total fluxes with caution, and perhaps refine the analysis with refined techniques, as done e.g. with the ``GHZ2'' object in Paper III. 

One can then estimate an aperture correction factor for each source from the detection image measurements, $q_{aper} = f_{tot,det} / f_{aper,det}$, and compute its total flux in each band as $f_{tot,band}= q_{aper} \times f_{aper,band}$. We verified on the simulated images that this procedure allows to obtain a good estimate of the colors, with typical median bias within 5\% of the input fluxes. 
Uncertainties are estimated in the same way, considering the aperture correction factor as a fixed parameter, and thus without propagating the errors on its measurement: $\sigma_{tot,band}= q_{aper} \times \sigma_{aper,band}$.
In the released catalog we provide the total fluxes computed using colors in apertures of 2 and 3 FWHM (0.28$\arcsec$, 0.42$\arcsec$ diameter, respectively), and in 0.1\arcsec radius.

\begin{figure*}
\center
\includegraphics[width=\textwidth]{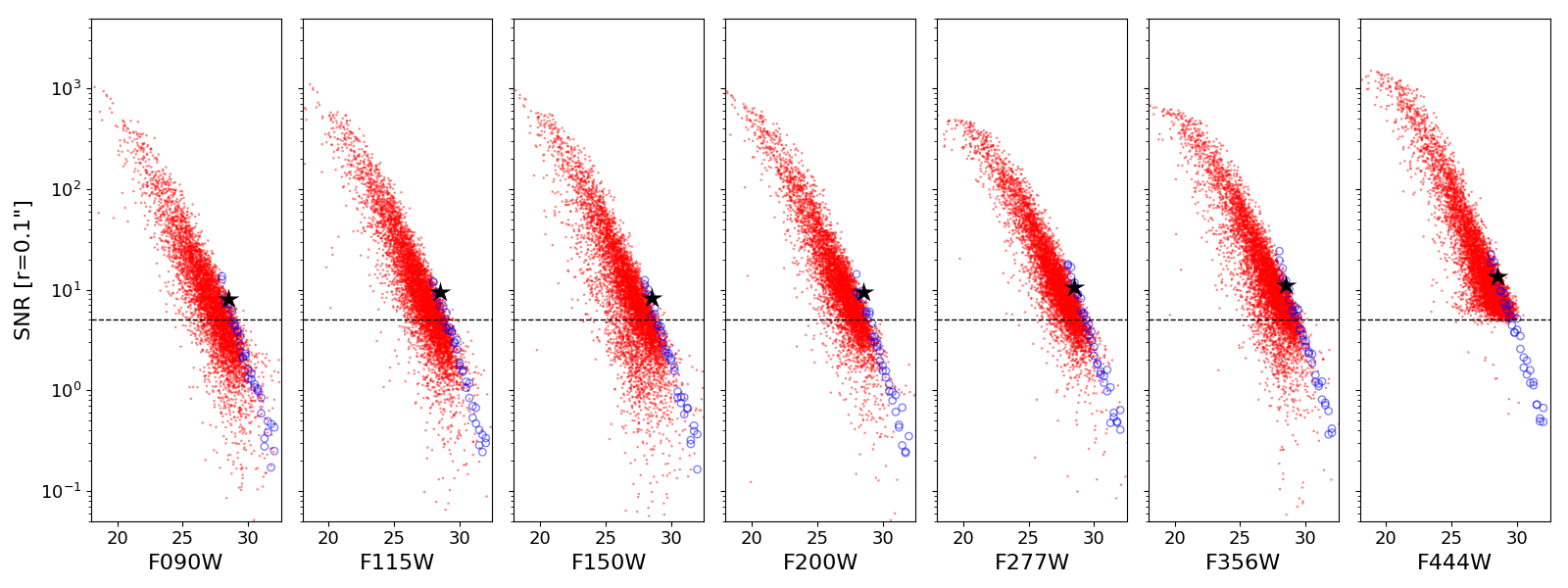}
 \caption{Signal-to-noise ratio in apertures of radius 0.1$\arcsec$, as a function of the total magnitude, in the seven bands. Red dots: all real detected sources; blue open circles: averages of the artificial point sources injected in the images to estimate depths (see text for details); black star: expected SNR for point sources at mag=28.5 (based on ETC estimates).}
 \label{snr}
\end{figure*}

The number counts of the detected sources as a function of their measured total magnitude in all bands, computed using the colors in 2 FWHM apertures on the PSF-matched images, are shown in the lower panel of Fig. \ref{counts}. The counts are given in arcmin$^{-2}$ using an estimated total area of 9.86 sq. arcmin for the F444W FoV. We also plot the number counts in \textit{Spitzer} IRAC-CH2 (4.5 $\mu$m) from the same area in the Hubble Frontier Fields images \citep{Lotz2017}, for comparison. 
In Fig. \ref{snr} we show the SNRs within apertures of radius 0.1$\arcsec$ of the detected sources, along with the SNR of the artificial point-sources used to estimate the depth of the images, as described in Sect. \ref{sec:data}. The SNR of real sources scatters toward lower values because they have different and typically less concentrated light profiles, resulting in varying and typically lower values. The 3$\sigma$-clipped median value for real F444W detections is 28.1 at 10$\sigma$.

\begin{figure}
\center
\includegraphics[width=0.48\textwidth]{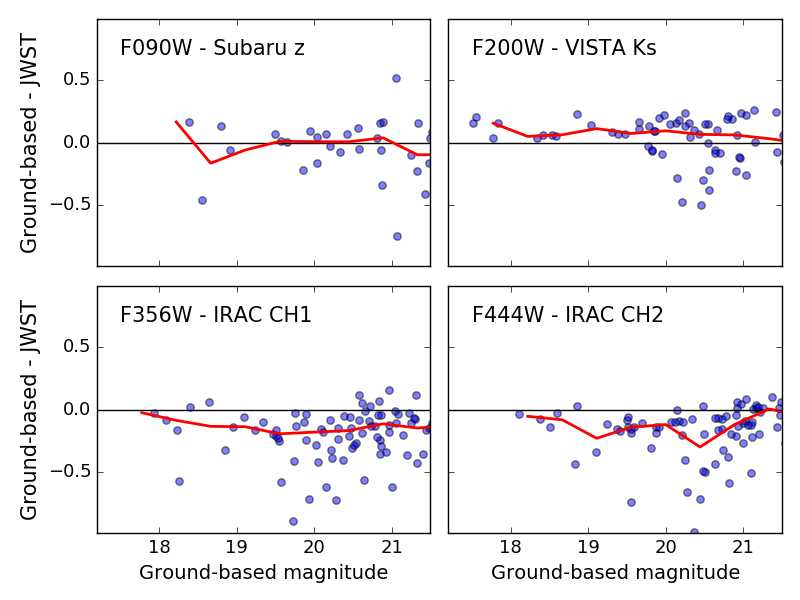}
 \caption{Comparison between NIRCam total magnitudes and catalogs obtained by running \textsc{SExtractor} on archival data, at four wavelengths sampled by similar filters by the JWST and the corresponding instrument (with a color correction applied on the basis of theoretical SEDs). The red lines are the medians of the distributions. The overall consistency is good and the typical average offsets are below 0.1-0.2 magnitudes.}
 \label{zp}
\end{figure}

As an external check, and to evaluate the reliability of the current photometric calibration (see Sect. \ref{sec:data}), we compared the magnitudes of the brightest galaxies in our catalog with those from  archival data. We considered four bands (Subaru $z$, VISTA $Ks$ and \textit{Spitzer} IRAC CH1 and CH2) with similar wavelength coverage and filter response curves to the NIRCam bands, adding a color correction factor computed by means of theoretical Spectral Energy Distributions  \citep[SEDs; using][models]{Bruzual2003}. The results of the test are shown in Fig. \ref{zp}, and confirm an overall good consistency, with median offsets lower than 0.1-0.2 magnitudes (red lines). 


\section{Caveats} \label{sec:caveats}

While the GLASS-ERS data-set is of exquisite quality, and the photometric techniques adopted in this work have been used and tested extensively on deep HST images in previous surveys, we are aware of some limitations that might affects our results, considering the limited knowledge of the instruments' actual capabilities.


\begin{itemize}
    \item As we mentioned several times, the processing of raw data was performed using the currently available versions of pipelines and reference files. We expect the results will  significantly improve when updated versions become available; 
    \item while the background light and many defects have been removed from the images, local variations and residual minor astrometric offsets are still present and can affect the photometry, especially of faint sources; 
    \item photometric calibration shall be refined, the currently estimated uncertainty being $\sim$0.1-0.2 magnitudes, based on cross-checks with external ground-based catalogs;
    \item we did not attempt to remove spurious detections (such as fragments of stellar spikes or pixel clumps at the borders of the mosaics) from the catalog; however we estimate they contribute to a small fraction of the total detections; 
    \item \textsc{WebbPSF} models are an approximate description of the real shape of the observed stars; 
    \item since aperture photometry is prone to contamination from neighboring sources, estimation of fluxes can be affected, especially in the redder bands where the density of objects is higher and they typically have larger dimensions. We mitigated the issue by masking the image using the segmentation map and choosing relatively small apertures to estimate colors, but a more thorough check should be performed. 
\end{itemize}

\section{Summary and conclusions}
\label{sec:conclusions}

We release the images and the corresponding photometric catalog of this very first JWST NIRCam deep extragalactic data, which are used in the accompanying first GLASS science papers. The images are obtained from the NIRISS parallel pointing of the GLASS-ERS Program; the data-set consists of mosaics in seven NIRCam bands (F090W, F115W, F150W, F200W, F277W, F356W) with typical depths of $\sim28.8-29.7$ at 5$\sigma$ (point sources). We processed them using a customised version of the STScI pipeline described in Section \ref{sec:data}, to remove defects in the raw exposures and obtain scientific mosaics aligned on a common grid of pixels with scale 0.031$\arcsec$. We release the images and the corresponding RMS maps. 

Using \textsc{SExtractor} we detected 6368 objects on the F444W image, and with \textsc{a-phot} we measured aperture photometry on PSF-matched versions of the images. The photometric calibration was checked against ground-based archival data resulting in an overall accuracy within 0.1-0.2 magnitudes. 
The released catalog contains coordinates, total fluxes and corresponding uncertainties of all the detected sources. Fluxes are obtained with the method described in Section \ref{sec:photometry}, i.e. using colors estimated on the PSF-matched images in circular apertures of diameters  0.28$\arcsec$ and 0.42$\arcsec$ (corresponding to 2 and 3 FWHM in F444W), and Kron apertures on the F444W band. The catalog also includes additional quantities of interest measured with \textsc{SExtractor} on the detection image, such as basic morphological parameters, flags and \texttt{CLASS\_STAR}. 

Images and catalogs are available for download from the GLASS-ERS collaboration website\footnote{\url{https://glass.astro.ucla.edu}} and from the \textsc{AstroDeep} website\footnote{\url{http://www.astrodeep.eu}}.

All the {\it JWST} data used in this paper can be found in MAST: \dataset[10.17909/fqaq-p393]{http://dx.doi.org/10.17909/fqaq-p393}.


\acknowledgments
This work is based on observations made with the NASA/ESA/CSA James Webb Space Telescope. The data were obtained from the Mikulski Archive for Space Telescopes at the Space Telescope Science Institute, which is operated by the Association of Universities for Research in Astronomy, Inc., under NASA contract NAS 5-03127 for JWST. These observations are associated with program JWST-ERS-1324. We acknowledge financial support from NASA through grant JWST-ERS-1324. This research is supported in part by the Australian Research Council Centre of Excellence for All Sky Astrophysics in 3 Dimensions (ASTRO 3D), through project number CE170100013. KG and TN acknowledge support from Australian Research Council Laureate Fellowship FL180100060. MB acknowledges support from the Slovenian national research agency ARRS through grant N1-0238. We acknowledge financial support through grants PRIN-MIUR 2017WSCC32 and 2020SKSTHZ. CM acknowledges support by the VILLUM FONDEN under grant 37459. The Cosmic Dawn Center (DAWN) is funded by the Danish National Research Foundation under grant DNRF140. This work has made use of data from the European Space Agency (ESA) mission 
{\it Gaia} (\url{https://www.cosmos.esa.int/gaia}), processed by the {\it Gaia}
Data Processing and Analysis Consortium (DPAC,
\url{https://www.cosmos.esa.int/web/gaia/dpac/consortium}). Funding for the DPAC
has been provided by national institutions, in particular the institutions
participating in the {\it Gaia} Multilateral Agreement. The authors thank Paola Marrese and Silvia Marinoni (Space Science Data Center, Italian Space Agency) for their contribution to the work.

\bibliography{biblio.bib}{}

\begin{thebibliography}{}
\expandafter\ifx\csname natexlab\endcsname\relax\def\natexlab#1{#1}\fi
\providecommand{\url}[1]{\href{#1}{#1}}
\providecommand{\dodoi}[1]{doi:~\href{http://doi.org/#1}{\nolinkurl{#1}}}
\providecommand{\doeprint}[1]{\href{http://ascl.net/#1}{\nolinkurl{http://ascl.net/#1}}}
\providecommand{\doarXiv}[1]{\href{https://arxiv.org/abs/#1}{\nolinkurl{https://arxiv.org/abs/#1}}}

\bibitem[{{Bertin}(2006)}]{Bertin2006}
{Bertin}, E. 2006, in Astronomical Society of the Pacific Conference Series,
  Vol. 351, Astronomical Data Analysis Software and Systems XV, ed.
  C.~{Gabriel}, C.~{Arviset}, D.~{Ponz}, \& E.~{Solano}, 112

\bibitem[{{Bertin} \& {Arnouts}(1996)}]{Bertin1996}
{Bertin}, E., \& {Arnouts}, S. 1996, \aaps, 117, 393

\bibitem[{{Bertin} {et~al.}(2002){Bertin}, {Mellier}, {Radovich}, {Missonnier},
  {Didelon}, \& {Morin}}]{Bertin2002}
{Bertin}, E., {Mellier}, Y., {Radovich}, M., {et~al.} 2002, in Astronomical
  Society of the Pacific Conference Series, Vol. 281, Astronomical Data
  Analysis Software and Systems XI, ed. D.~A. {Bohlender}, D.~{Durand}, \&
  T.~H. {Handley}, 228

\bibitem[{{Boucaud} {et~al.}(2016){Boucaud}, {Bocchio}, {Abergel}, {Orieux},
  {Dole}, \& {Hadj-Youcef}}]{Boucaud2016}
{Boucaud}, A., {Bocchio}, M., {Abergel}, A., {et~al.} 2016, \aap, 596, A63,
  \dodoi{10.1051/0004-6361/201629080}

\bibitem[{{Bruzual} \& {Charlot}(2003)}]{Bruzual2003}
{Bruzual}, G., \& {Charlot}, S. 2003, \mnras, 344, 1000,
  \dodoi{10.1046/j.1365-8711.2003.06897.x}

\bibitem[{{Burriesci}(2005)}]{Burriesci2005}
{Burriesci}, L.~G. 2005, in Society of Photo-Optical Instrumentation Engineers
  (SPIE) Conference Series, Vol. 5904, Cryogenic Optical Systems and
  Instruments XI, ed. J.~B. {Heaney} \& L.~G. {Burriesci}, 21--29,
  \dodoi{10.1117/12.613596}

\bibitem[{{Castellano} {et~al.}(2016){Castellano}, {Amor{\'{\i}}n}, {Merlin},
  {Fontana}, {McLure}, {M{\'a}rmol-Queralt{\'o}}, {Mortlock}, {Parsa},
  {Dunlop}, {Elbaz}, {Balestra}, {Boucaud}, {Bourne}, {Boutsia}, {Brammer},
  {Bruce}, {Buitrago}, {Capak}, {Cappelluti}, {Ciesla}, {Comastri}, {Cullen},
  {Derriere}, {Faber}, {Giallongo}, {Grazian}, {Grillo}, {Mercurio},
  {Micha{\l}owski}, {Nonino}, {Paris}, {Pentericci}, {Pilo}, {Rosati},
  {Santini}, {Schreiber}, {Shu}, \& {Wang}}]{Castellano2016}
{Castellano}, M., {Amor{\'{\i}}n}, R., {Merlin}, E., {et~al.} 2016, \aap, 590,
  A31, \dodoi{10.1051/0004-6361/201527514}

\bibitem[{{Castellano} {et~al.}(2022){Castellano}, {Fontana}, {Treu},
  {Santini}, {Merlin}, {Leethochawalit}, {Trenti}, {Mestric}, {Vanzella},
  {Bonchi}, {Belfiori}, {Nonino}, {Paris}, {Polenta}, {Roberts-Borsani},
  {Boyett}, {Calabro}, {Glazebrook}, {Grillo}, {Mascia}, {Mason}, {Mercurio},
  {Morishita}, {Nanayakkara}, {Pentericci}, {Rosati}, {Vulcani}, {Wang}, \&
  {Yang}}]{Castellano2022}
{Castellano}, M., {Fontana}, A., {Treu}, T., {et~al.} 2022, arXiv e-prints,
  arXiv:2207.09436.
\newblock \doarXiv{2207.09436}

\bibitem[{{Ding} {et~al.}(2020){Ding}, {Silverman}, {Treu}, {Schulze},
  {Schramm}, {Birrer}, {Park}, {Jahnke}, {Bennert}, {Kartaltepe}, {Koekemoer},
  {Malkan}, \& {Sanders}}]{Ding2020}
{Ding}, X., {Silverman}, J., {Treu}, T., {et~al.} 2020, \apj, 888, 37,
  \dodoi{10.3847/1538-4357/ab5b90}

\bibitem[{{Gaia Collaboration} {et~al.}(2016){Gaia Collaboration}, {Prusti},
  {de Bruijne}, {Brown}, {Vallenari}, {Babusiaux}, {Bailer-Jones}, {Bastian},
  {Biermann}, {Evans}, {Eyer}, {Jansen}, {Jordi}, {Klioner}, {Lammers},
  {Lindegren}, {Luri}, {Mignard}, {Milligan}, {Panem}, {Poinsignon},
  {Pourbaix}, {Randich}, {Sarri}, {Sartoretti}, {Siddiqui}, {Soubiran},
  {Valette}, {van Leeuwen}, {Walton}, {Aerts}, {Arenou}, {Cropper}, {Drimmel},
  {H{\o}g}, {Katz}, {Lattanzi}, {O'Mullane}, {Grebel}, {Holland}, {Huc},
  {Passot}, {Bramante}, {Cacciari}, {Casta{\~n}eda}, {Chaoul}, {Cheek}, {De
  Angeli}, {Fabricius}, {Guerra}, {Hern{\'a}ndez}, {Jean-Antoine-Piccolo},
  {Masana}, {Messineo}, {Mowlavi}, {Nienartowicz}, {Ord{\'o}{\~n}ez-Blanco},
  {Panuzzo}, {Portell}, {Richards}, {Riello}, {Seabroke}, {Tanga},
  {Th{\'e}venin}, {Torra}, {Els}, {Gracia-Abril}, {Comoretto},
  {Garcia-Reinaldos}, {Lock}, {Mercier}, {Altmann}, {Andrae}, {Astraatmadja},
  {Bellas-Velidis}, {Benson}, {Berthier}, {Blomme}, {Busso}, {Carry},
  {Cellino}, {Clementini}, {Cowell}, {Creevey}, {Cuypers}, {Davidson}, {De
  Ridder}, {de Torres}, {Delchambre}, {Dell'Oro}, {Ducourant}, {Fr{\'e}mat},
  {Garc{\'\i}a-Torres}, {Gosset}, {Halbwachs}, {Hambly}, {Harrison}, {Hauser},
  {Hestroffer}, {Hodgkin}, {Huckle}, {Hutton}, {Jasniewicz}, {Jordan},
  {Kontizas}, {Korn}, {Lanzafame}, {Manteiga}, {Moitinho}, {Muinonen},
  {Osinde}, {Pancino}, {Pauwels}, {Petit}, {Recio-Blanco}, {Robin}, {Sarro},
  {Siopis}, {Smith}, {Smith}, {Sozzetti}, {Thuillot}, {van Reeven}, {Viala},
  {Abbas}, {Abreu Aramburu}, {Accart}, {Aguado}, {Allan}, {Allasia},
  {Altavilla}, {{\'A}lvarez}, {Alves}, {Anderson}, {Andrei}, {Anglada Varela},
  {Antiche}, {Antoja}, {Ant{\'o}n}, {Arcay}, {Atzei}, {Ayache}, {Bach},
  {Baker}, {Balaguer-N{\'u}{\~n}ez}, {Barache}, {Barata}, {Barbier}, {Barblan},
  {Baroni}, {Barrado y Navascu{\'e}s}, {Barros}, {Barstow}, {Becciani},
  {Bellazzini}, {Bellei}, {Bello Garc{\'\i}a}, {Belokurov}, {Bendjoya},
  {Berihuete}, {Bianchi}, {Bienaym{\'e}}, {Billebaud}, {Blagorodnova},
  {Blanco-Cuaresma}, {Boch}, {Bombrun}, {Borrachero}, {Bouquillon}, {Bourda},
  {Bouy}, {Bragaglia}, {Breddels}, {Brouillet}, {Br{\"u}semeister},
  {Bucciarelli}, {Budnik}, {Burgess}, {Burgon}, {Burlacu}, {Busonero}, {Buzzi},
  {Caffau}, {Cambras}, {Campbell}, {Cancelliere}, {Cantat-Gaudin}, {Carlucci},
  {Carrasco}, {Castellani}, {Charlot}, {Charnas}, {Charvet}, {Chassat},
  {Chiavassa}, {Clotet}, {Cocozza}, {Collins}, {Collins}, {Costigan}, {Crifo},
  {Cross}, {Crosta}, {Crowley}, {Dafonte}, {Damerdji}, {Dapergolas}, {David},
  {David}, {De Cat}, {de Felice}, {de Laverny}, {De Luise}, {De March}, {de
  Martino}, {de Souza}, {Debosscher}, {del Pozo}, {Delbo}, {Delgado},
  {Delgado}, {di Marco}, {Di Matteo}, {Diakite}, {Distefano}, {Dolding}, {Dos
  Anjos}, {Drazinos}, {Dur{\'a}n}, {Dzigan}, {Ecale}, {Edvardsson}, {Enke},
  {Erdmann}, {Escolar}, {Espina}, {Evans}, {Eynard Bontemps}, {Fabre},
  {Fabrizio}, {Faigler}, {Falc{\~a}o}, {Farr{\`a}s Casas}, {Faye}, {Federici},
  {Fedorets}, {Fern{\'a}ndez-Hern{\'a}ndez}, {Fernique}, {Fienga}, {Figueras},
  {Filippi}, {Findeisen}, {Fonti}, {Fouesneau}, {Fraile}, {Fraser}, {Fuchs},
  {Furnell}, {Gai}, {Galleti}, {Galluccio}, {Garabato}, {Garc{\'\i}a-Sedano},
  {Gar{\'e}}, {Garofalo}, {Garralda}, {Gavras}, {Gerssen}, {Geyer}, {Gilmore},
  {Girona}, {Giuffrida}, {Gomes}, {Gonz{\'a}lez-Marcos},
  {Gonz{\'a}lez-N{\'u}{\~n}ez}, {Gonz{\'a}lez-Vidal}, {Granvik}, {Guerrier},
  {Guillout}, {Guiraud}, {G{\'u}rpide}, {Guti{\'e}rrez-S{\'a}nchez}, {Guy},
  {Haigron}, {Hatzidimitriou}, {Haywood}, {Heiter}, {Helmi}, {Hobbs},
  {Hofmann}, {Holl}, {Holland}, {Hunt}, {Hypki}, {Icardi}, {Irwin}, {Jevardat
  de Fombelle}, {Jofr{\'e}}, {Jonker}, {Jorissen}, {Julbe}, {Karampelas},
  {Kochoska}, {Kohley}, {Kolenberg}, {Kontizas}, {Koposov}, {Kordopatis},
  {Koubsky}, {Kowalczyk}, {Krone-Martins}, {Kudryashova}, {Kull}, {Bachchan},
  {Lacoste-Seris}, {Lanza}, {Lavigne}, {Le Poncin-Lafitte}, {Lebreton},
  {Lebzelter}, {Leccia}, {Leclerc}, {Lecoeur-Taibi}, {Lemaitre}, {Lenhardt},
  {Leroux}, {Liao}, {Licata}, {Lindstr{\o}m}, {Lister}, {Livanou}, {Lobel},
  {L{\"o}ffler}, {L{\'o}pez}, {Lopez-Lozano}, {Lorenz}, {Loureiro},
  {MacDonald}, {Magalh{\~a}es Fernandes}, {Managau}, {Mann}, {Mantelet},
  {Marchal}, {Marchant}, {Marconi}, {Marie}, {Marinoni}, {Marrese},
  {Marschalk{\'o}}, {Marshall}, {Mart{\'\i}n-Fleitas}, {Martino}, {Mary},
  {Matijevi{\v{c}}}, {Mazeh}, {McMillan}, {Messina}, {Mestre}, {Michalik},
  {Millar}, {Miranda}, {Molina}, {Molinaro}, {Molinaro}, {Moln{\'a}r},
  {Moniez}, {Montegriffo}, {Monteiro}, {Mor}, {Mora}, {Morbidelli}, {Morel},
  {Morgenthaler}, {Morley}, {Morris}, {Mulone}, {Muraveva}, {Musella},
  {Narbonne}, {Nelemans}, {Nicastro}, {Noval}, {Ord{\'e}novic},
  {Ordieres-Mer{\'e}}, {Osborne}, {Pagani}, {Pagano}, {Pailler}, {Palacin},
  {Palaversa}, {Parsons}, {Paulsen}, {Pecoraro}, {Pedrosa}, {Pentik{\"a}inen},
  {Pereira}, {Pichon}, {Piersimoni}, {Pineau}, {Plachy}, {Plum}, {Poujoulet},
  {Pr{\v{s}}a}, {Pulone}, {Ragaini}, {Rago}, {Rambaux}, {Ramos-Lerate},
  {Ranalli}, {Rauw}, {Read}, {Regibo}, {Renk}, {Reyl{\'e}}, {Ribeiro},
  {Rimoldini}, {Ripepi}, {Riva}, {Rixon}, {Roelens}, {Romero-G{\'o}mez},
  {Rowell}, {Royer}, {Rudolph}, {Ruiz-Dern}, {Sadowski}, {Sagrist{\`a}
  Sell{\'e}s}, {Sahlmann}, {Salgado}, {Salguero}, {Sarasso}, {Savietto},
  {Schnorhk}, {Schultheis}, {Sciacca}, {Segol}, {Segovia}, {Segransan},
  {Serpell}, {Shih}, {Smareglia}, {Smart}, {Smith}, {Solano}, {Solitro},
  {Sordo}, {Soria Nieto}, {Souchay}, {Spagna}, {Spoto}, {Stampa}, {Steele},
  {Steidelm{\"u}ller}, {Stephenson}, {Stoev}, {Suess}, {S{\"u}veges}, {Surdej},
  {Szabados}, {Szegedi-Elek}, {Tapiador}, {Taris}, {Tauran}, {Taylor},
  {Teixeira}, {Terrett}, {Tingley}, {Trager}, {Turon}, {Ulla}, {Utrilla},
  {Valentini}, {van Elteren}, {Van Hemelryck}, {van Leeuwen}, {Varadi},
  {Vecchiato}, {Veljanoski}, {Via}, {Vicente}, {Vogt}, {Voss}, {Votruba},
  {Voutsinas}, {Walmsley}, {Weiler}, {Weingrill}, {Werner}, {Wevers},
  {Whitehead}, {Wyrzykowski}, {Yoldas}, {{\v{Z}}erjal}, {Zucker}, {Zurbach},
  {Zwitter}, {Alecu}, {Allen}, {Allende Prieto}, {Amorim},
  {Anglada-Escud{\'e}}, {Arsenijevic}, {Azaz}, {Balm}, {Beck}, {Bernstein},
  {Bigot}, {Bijaoui}, {Blasco}, {Bonfigli}, {Bono}, {Boudreault}, {Bressan},
  {Brown}, {Brunet}, {Bunclark}, {Buonanno}, {Butkevich}, {Carret}, {Carrion},
  {Chemin}, {Ch{\'e}reau}, {Corcione}, {Darmigny}, {de Boer}, {de Teodoro}, {de
  Zeeuw}, {Delle Luche}, {Domingues}, {Dubath}, {Fodor}, {Fr{\'e}zouls},
  {Fries}, {Fustes}, {Fyfe}, {Gallardo}, {Gallegos}, {Gardiol}, {Gebran},
  {Gomboc}, {G{\'o}mez}, {Grux}, {Gueguen}, {Heyrovsky}, {Hoar}, {Iannicola},
  {Isasi Parache}, {Janotto}, {Joliet}, {Jonckheere}, {Keil}, {Kim},
  {Klagyivik}, {Klar}, {Knude}, {Kochukhov}, {Kolka}, {Kos}, {Kutka}, {Lainey},
  {LeBouquin}, {Liu}, {Loreggia}, {Makarov}, {Marseille}, {Martayan},
  {Martinez-Rubi}, {Massart}, {Meynadier}, {Mignot}, {Munari}, {Nguyen},
  {Nordlander}, {Ocvirk}, {O'Flaherty}, {Olias Sanz}, {Ortiz}, {Osorio},
  {Oszkiewicz}, {Ouzounis}, {Palmer}, {Park}, {Pasquato}, {Peltzer}, {Peralta},
  {P{\'e}turaud}, {Pieniluoma}, {Pigozzi}, {Poels}, {Prat}, {Prod'homme},
  {Raison}, {Rebordao}, {Risquez}, {Rocca-Volmerange}, {Rosen}, {Ruiz-Fuertes},
  {Russo}, {Sembay}, {Serraller Vizcaino}, {Short}, {Siebert}, {Silva},
  {Sinachopoulos}, {Slezak}, {Soffel}, {Sosnowska}, {Strai{\v{z}}ys}, {ter
  Linden}, {Terrell}, {Theil}, {Tiede}, {Troisi}, {Tsalmantza}, {Tur},
  {Vaccari}, {Vachier}, {Valles}, {Van Hamme}, {Veltz}, {Virtanen}, {Wallut},
  {Wichmann}, {Wilkinson}, {Ziaeepour}, \& {Zschocke}}]{Prusti2016}
{Gaia Collaboration}, {Prusti}, T., {de Bruijne}, J.~H.~J., {et~al.} 2016,
  \aap, 595, A1, \dodoi{10.1051/0004-6361/201629272}

\bibitem[{{Galametz} {et~al.}(2013){Galametz}, {Grazian}, {Fontana},
  {Ferguson}, {Ashby}, {Barro}, {Castellano}, {Dahlen}, {Donley}, {Faber},
  {Grogin}, {Guo}, {Huang}, {Kocevski}, {Koekemoer}, {Lee}, {McGrath}, {Peth},
  {Willner}, {Almaini}, {Cooper}, {Cooray}, {Conselice}, {Dickinson}, {Dunlop},
  {Fazio}, {Foucaud}, {Gardner}, {Giavalisco}, {Hathi}, {Hartley}, {Koo},
  {Lai}, {de Mello}, {McLure}, {Lucas}, {Paris}, {Pentericci}, {Santini},
  {Simpson}, {Sommariva}, {Targett}, {Weiner}, {Wuyts}, \& {the CANDELS
  Team}}]{Galametz2013}
{Galametz}, A., {Grazian}, A., {Fontana}, A., {et~al.} 2013, \apjs, 206, 10,
  \dodoi{10.1088/0067-0049/206/2/10}

\bibitem[{{Girardi} {et~al.}(2005){Girardi}, {Groenewegen}, {Hatziminaoglou},
  \& {da Costa}}]{Girardi2005}
{Girardi}, L., {Groenewegen}, M.~A.~T., {Hatziminaoglou}, E., \& {da Costa}, L.
  2005, \aap, 436, 895, \dodoi{10.1051/0004-6361:20042352}

\bibitem[{{Girardi} {et~al.}(2012){Girardi}, {Barbieri}, {Groenewegen},
  {Marigo}, {Bressan}, {Rocha-Pinto}, {Santiago}, {Camargo}, \& {da
  Costa}}]{Girardi2012}
{Girardi}, L., {Barbieri}, M., {Groenewegen}, M. A.~T., {et~al.} 2012, in
  Astrophysics and Space Science Proceedings, Vol.~26, Red Giants as Probes of
  the Structure and Evolution of the Milky Way, 165,
  \dodoi{10.1007/978-3-642-18418-5\_17}

\bibitem[{{Grogin} {et~al.}(2011){Grogin}, {Kocevski}, {Faber}, {Ferguson},
  {Koekemoer}, {Riess}, {Acquaviva}, {Alexander}, {Almaini}, {Ashby}, {Barden},
  {Bell}, {Bournaud}, {Brown}, {Caputi}, {Casertano}, {Cassata}, {Castellano},
  {Challis}, {Chary}, {Cheung}, {Cirasuolo}, {Conselice}, {Roshan Cooray},
  {Croton}, {Daddi}, {Dahlen}, {Dav{\'e}}, {de Mello}, {Dekel}, {Dickinson},
  {Dolch}, {Donley}, {Dunlop}, {Dutton}, {Elbaz}, {Fazio}, {Filippenko},
  {Finkelstein}, {Fontana}, {Gardner}, {Garnavich}, {Gawiser}, {Giavalisco},
  {Grazian}, {Guo}, {Hathi}, {H{\"a}ussler}, {Hopkins}, {Huang}, {Huang},
  {Jha}, {Kartaltepe}, {Kirshner}, {Koo}, {Lai}, {Lee}, {Li}, {Lotz}, {Lucas},
  {Madau}, {McCarthy}, {McGrath}, {McIntosh}, {McLure}, {Mobasher},
  {Moustakas}, {Mozena}, {Nandra}, {Newman}, {Niemi}, {Noeske}, {Papovich},
  {Pentericci}, {Pope}, {Primack}, {Rajan}, {Ravindranath}, {Reddy}, {Renzini},
  {Rix}, {Robaina}, {Rodney}, {Rosario}, {Rosati}, {Salimbeni}, {Scarlata},
  {Siana}, {Simard}, {Smidt}, {Somerville}, {Spinrad}, {Straughn}, {Strolger},
  {Telford}, {Teplitz}, {Trump}, {van der Wel}, {Villforth}, {Wechsler},
  {Weiner}, {Wiklind}, {Wild}, {Wilson}, {Wuyts}, {Yan}, \& {Yun}}]{Grogin2011}
{Grogin}, N.~A., {Kocevski}, D.~D., {Faber}, S.~M., {et~al.} 2011, 197, 35,
  \dodoi{10.1088/0067-0049/197/2/35}

\bibitem[{{Koekemoer} {et~al.}(2011){Koekemoer}, {Faber}, {Ferguson}, {Grogin},
  {Kocevski}, {Koo}, {Lai}, {Lotz}, {Lucas}, {McGrath}, {Ogaz}, {Rajan},
  {Riess}, {Rodney}, {Strolger}, {Casertano}, {Castellano}, {Dahlen},
  {Dickinson}, {Dolch}, {Fontana}, {Giavalisco}, {Grazian}, {Guo}, {Hathi},
  {Huang}, {van der Wel}, {Yan}, {Acquaviva}, {Alexander}, {Almaini}, {Ashby},
  {Barden}, {Bell}, {Bournaud}, {Brown}, {Caputi}, {Cassata}, {Challis},
  {Chary}, {Cheung}, {Cirasuolo}, {Conselice}, {Roshan Cooray}, {Croton},
  {Daddi}, {Dav{\'e}}, {de Mello}, {de Ravel}, {Dekel}, {Donley}, {Dunlop},
  {Dutton}, {Elbaz}, {Fazio}, {Filippenko}, {Finkelstein}, {Frazer}, {Gardner},
  {Garnavich}, {Gawiser}, {Gruetzbauch}, {Hartley}, {H{\"a}ussler},
  {Herrington}, {Hopkins}, {Huang}, {Jha}, {Johnson}, {Kartaltepe},
  {Khostovan}, {Kirshner}, {Lani}, {Lee}, {Li}, {Madau}, {McCarthy},
  {McIntosh}, {McLure}, {McPartland}, {Mobasher}, {Moreira}, {Mortlock},
  {Moustakas}, {Mozena}, {Nandra}, {Newman}, {Nielsen}, {Niemi}, {Noeske},
  {Papovich}, {Pentericci}, {Pope}, {Primack}, {Ravindranath}, {Reddy},
  {Renzini}, {Rix}, {Robaina}, {Rosario}, {Rosati}, {Salimbeni}, {Scarlata},
  {Siana}, {Simard}, {Smidt}, {Snyder}, {Somerville}, {Spinrad}, {Straughn},
  {Telford}, {Teplitz}, {Trump}, {Vargas}, {Villforth}, {Wagner}, {Wandro},
  {Wechsler}, {Weiner}, {Wiklind}, {Wild}, {Wilson}, {Wuyts}, \&
  {Yun}}]{Koekemoer2011}
{Koekemoer}, A.~M., {Faber}, S.~M., {Ferguson}, H.~C., {et~al.} 2011, \apjs,
  197, 36, \dodoi{10.1088/0067-0049/197/2/36}

\bibitem[{{Kron}(1980)}]{Kron1980}
{Kron}, R.~G. 1980, \apjs, 43, 305, \dodoi{10.1086/190669}

\bibitem[{{Lotz} {et~al.}(2017){Lotz}, {Koekemoer}, {Coe}, {Grogin}, {Capak},
  {Mack}, {Anderson}, {Avila}, {Barker}, {Borncamp}, {Brammer}, {Durbin},
  {Gunning}, {Hilbert}, {Jenkner}, {Khandrika}, {Levay}, {Lucas}, {MacKenty},
  {Ogaz}, {Porterfield}, {Reid}, {Robberto}, {Royle}, {Smith},
  {Storrie-Lombardi}, {Sunnquist}, {Surace}, {Taylor}, {Williams}, {Bullock},
  {Dickinson}, {Finkelstein}, {Natarajan}, {Richard}, {Robertson}, {Tumlinson},
  {Zitrin}, {Flanagan}, {Sembach}, {Soifer}, \& {Mountain}}]{Lotz2017}
{Lotz}, J.~M., {Koekemoer}, A., {Coe}, D., {et~al.} 2017, \apj, 837, 97,
  \dodoi{10.3847/1538-4357/837/1/97}

\bibitem[{{Merlin} {et~al.}(2015){Merlin}, {Fontana}, {Ferguson}, {Dunlop},
  {Elbaz}, {Bourne}, {Bruce}, {Buitrago}, {Castellano}, {Schreiber}, {Grazian},
  {McLure}, {Okumura}, {Shu}, {Wang}, {Amor{\'{\i}}n}, {Boutsia}, {Cappelluti},
  {Comastri}, {Derriere}, {Faber}, \& {Santini}}]{Merlin2015}
{Merlin}, E., {Fontana}, A., {Ferguson}, H.~C., {et~al.} 2015, \aap, 582, A15,
  \dodoi{10.1051/0004-6361/201526471}

\bibitem[{{Merlin} {et~al.}(2016{\natexlab{a}}){Merlin}, {Amor{\'\i}n},
  {Castellano}, {Fontana}, {Buitrago}, {Dunlop}, {Elbaz}, {Boucaud}, {Bourne},
  {Boutsia}, {Brammer}, {Bruce}, {Capak}, {Cappelluti}, {Ciesla}, {Comastri},
  {Cullen}, {Derriere}, {Faber}, {Ferguson}, {Giallongo}, {Grazian}, {Lotz},
  {Micha{\l}owski}, {Paris}, {Pentericci}, {Pilo}, {Santini}, {Schreiber},
  {Shu}, \& {Wang}}]{Merlin2016b}
{Merlin}, E., {Amor{\'\i}n}, R., {Castellano}, M., {et~al.} 2016{\natexlab{a}},
  \aap, 590, A30, \dodoi{10.1051/0004-6361/201527513}

\bibitem[{{Merlin} {et~al.}(2016{\natexlab{b}}){Merlin}, {Bourne},
  {Castellano}, {Ferguson}, {Wang}, {Derriere}, {Dunlop}, {Elbaz}, \&
  {Fontana}}]{Merlin2016a}
{Merlin}, E., {Bourne}, N., {Castellano}, M., {et~al.} 2016{\natexlab{b}},
  \aap, 595, A97, \dodoi{10.1051/0004-6361/201628751}

\bibitem[{{Merlin} {et~al.}(2019){Merlin}, {Fortuni}, {Torelli}, {Santini},
  {Castellano}, {Fontana}, {Grazian}, {Pentericci}, {Pilo}, \&
  {Schmidt}}]{Merlin2019}
{Merlin}, E., {Fortuni}, F., {Torelli}, M., {et~al.} 2019, \mnras, 490, 3309,
  \dodoi{10.1093/mnras/stz2615}

\bibitem[{{Merlin} {et~al.}(2021){Merlin}, {Castellano}, {Santini},
  {Cipolletta}, {Boutsia}, {Schreiber}, {Buitrago}, {Fontana}, {Elbaz},
  {Dunlop}, {Grazian}, {McLure}, {McLeod}, {Nonino}, {Milvang-Jensen},
  {Derriere}, {Hathi}, {Pentericci}, {Fortuni}, \& {Calabr{\`o}}}]{Merlin2021}
{Merlin}, E., {Castellano}, M., {Santini}, P., {et~al.} 2021, \aap, 649, A22,
  \dodoi{10.1051/0004-6361/202140310}

\bibitem[{{Oke} \& {Gunn}(1983)}]{oke83}
{Oke}, J.~B., \& {Gunn}, J.~E. 1983, \apj, 266, 713, \dodoi{10.1086/160817}

\bibitem[{{Rieke} {et~al.}(2005){Rieke}, {Kelly}, {Horner}, \& {NIRCam
  Team}}]{Rieke2005}
{Rieke}, M., {Kelly}, D., {Horner}, S., \& {NIRCam Team}. 2005, in American
  Astronomical Society Meeting Abstracts, Vol. 207, American Astronomical
  Society Meeting Abstracts, 115.09

\bibitem[{{Rigby} {et~al.}(2022){Rigby}, {Perrin}, {McElwain}, {Kimble},
  {Friedman}, {Lallo}, {Doyon}, {Feinberg}, {Ferruit}, {Glasse}, {Rieke},
  {Rieke}, {Wright}, {Willott}, {Colon}, {Milam}, {Neff}, {Stark}, {Valenti},
  {Abell}, {Abney}, {Abul-Huda}, {Acton}, {Adams}, {Adler}, {Aguilar}, {Ahmed},
  {Albert}, {Alberts}, {Aldridge}, {Allen}, {Altenburg}, {Alves de Oliveira},
  {Anderson}, {Anderson}, {Anderson}, {Argyriou}, {Armstrong}, {Arribas},
  {Artigau}, {Arvai}, {Atkinson}, {Bacon}, {Bair}, {Banks}, {Barrientes},
  {Barringer}, {Bartosik}, {Bast}, {Baudoz}, {Beatty}, {Bechtold}, {Beck},
  {Bergeron}, {Bergkoetter}, {Bhatawdekar}, {Birkmann}, {Blazek}, {Blome},
  {Boccaletti}, {Boeker}, {Boia}, {Bonaventura}, {Bond}, {Bosley}, {Boucarut},
  {Bourque}, {Bouwman}, {Bower}, {Bowers}, {Boyer}, {Brady}, {Braun}, {Breda},
  {Bresnahan}, {Bright}, {Britt}, {Bromenschenkel}, {Brooks}, {Brooks},
  {Brown}, {Brown}, {Brown}, {Bunker}, {Burger}, {Bushouse}, {Cale}, {Cameron},
  {Cameron}, {Canipe}, {Caplinger}, {Caputo}, {Carey}, {Carniani},
  {Carrasquilla}, {Carruthers}, {Case}, {Chance}, {Chapman}, {Charlot},
  {Charlow}, {Chayer}, {Chen}, {Cherinka}, {Chichester}, {Chilton}, {Chonis},
  {Clark}, {Clark}, {Coe}, {Coleman}, {Comber}, {Comeau}, {Connolly}, {Cooper},
  {Cooper}, {Coppock}, {Correnti}, {Cossou}, {Coulais}, {Coyle}, {Cracraft},
  {Curti}, {Cuturic}, {Davis}, {Davis}, {Dean}, {DeLisa}, {deMeester},
  {Dencheva}, {Dencheva}, {DePasquale}, {Deschenes}, {Hunor Detre}, {Diaz},
  {Dicken}, {DiFelice}, {Dillman}, {Dixon}, {Doggett}, {Donaldson}, {Douglas},
  {DuPrie}, {Dupuis}, {Durning}, {Easmin}, {Eck}, {Edeani}, {Egami},
  {Ehrenwinkler}, {Eisenhamer}, {Eisenhower}, {Elie}, {Elliott}, {Elliott},
  {Ellis}, {Engesser}, {Espinoza}, {Etienne}, {Etxaluze}, {Falini}, {Feeney},
  {Ferry}, {Filippazzo}, {Fincham}, {Fix}, {Flagey}, {Florian}, {Flynn},
  {Fontanella}, {Ford}, {Forshay}, {Fox}, {Franz}, {Fu}, {Fullerton}, {Galkin},
  {Galyer}, {Garcia Marin}, {Gardner}, {Gardner}, {Garland}, {Gasman},
  {Gaspar}, {Gaudreau}, {Gauthier}, {Geers}, {Geithner}, {Gennaro}, {Giardino},
  {Girard}, {Giuliano}, {Glassmire}, {Glauser}, {Glazer}, {Godfrey},
  {Golimowski}, {Gollnitz}, {Gong}, {Gonzaga}, {Gordon}, {Gordon},
  {Goudfrooij}, {Greene}, {Greenhouse}, {Grimaldi}, {Groebner}, {Grundy},
  {Guillard}, {Gutman}, {Ha}, {Haderlein}, {Hagedorn}, {Hainline}, {Haley},
  {Hami}, {Hamilton}, {Hammel}, {Hansen}, {Harkins}, {Harr}, {Hart}, {Hart},
  {Hartig}, {Hashimoto}, {Haskins}, {Hathaway}, {Havey}, {Hayden}, {Hecht},
  {Heller-Boyer}, {Henry}, {Hermann}, {Hernandez}, {Hesman}, {Hicks},
  {Hilbert}, {Hines}, {Hoffman}, {Holfeltz}, {Holler}, {Hoppa}, {Hott},
  {Howard}, {Hunter}, {Hunter}, {Hurst}, {Husemann}, {Hustak}, {Ilinca Ignat},
  {Irish}, {Jackson}, {Jahromi}, {Jakobsen}, {James}, {James}, {Januszewski},
  {Jenkins}, {Jirdeh}, {Johnson}, {Johnson}, {Jones}, {Jones}, {Jones},
  {Jones}, {Jordan}, {Jordan}, {Jurczyk}, {Jurling}, {Kaleida}, {Kalmanson},
  {Kammerer}, {Kang}, {Kao}, {Karakla}, {Kavanagh}, {Kelly}, {Kendrew},
  {Kennedy}, {Kenny}, {Keski-kuha}, {Keyes}, {Kidwell}, {Kinzel}, {Kirk},
  {Kirkpatrick}, {Kirshenblat}, {Klaassen}, {Knapp}, {Knight}, {Knollenberg},
  {Koehler}, {Koekemoer}, {Kovacs}, {Kulp}, {Kumari}, {Kyprianou}, {La Massa},
  {Labador}, {Labiano Ortega}, {Lagage}, {Lajoie}, {Lallo}, {Lam}, {Lamb},
  {Lambros}, {Lampenfield}, {Langston}, {Larson}, {Law}, {Lawrence}, {Lee},
  {Leisenring}, {Lepo}, {Leveille}, {Levenson}, {Levine}, {Levy}, {Lewis},
  {Lewis}, {Libralato}, {Lightsey}, {Link}, {Liu}, {Lo}, {Lockwood}, {Logue},
  {Long}, {Long}, {Loomis}, {Lopez-Caniego}, {Alvarez}, {Love-Pruitt}, {Lucy},
  {Luetzgendorf}, {Maghami}, {Maiolino}, {Major}, {Malla}, {Malumuth},
  {Manjavacas}, {Mannfolk}, {Marrione}, {Marston}, {Martel}, {Maschmann},
  {Masci}, {Masciarelli}, {Maszkiewicz}, {Mather}, {McKenzie}, {McLean},
  {McMaster}, {Melbourne}, {Mel{\'e}ndez}, {Menzel}, {Merz}, {Meyett}, {Meza},
  {Miskey}, {Misselt}, {Moller}, {Morrison}, {Morse}, {Moseley}, {Mosier},
  {Mountain}, {Mueckay}, {Mueller}, {Mullally}, {Murphy}, {Murray}, {Murray},
  {Muzerolle}, {Mycroft}, {Myers}, {Myrick}, {Nanavati}, {Nance}, {Nayak},
  {Naylor}, {Nelan}, {Nickson}, {Nielson}, {Nieto-Santisteban}, {Nikolov},
  {Noriega-Crespo}, {O'Shaughnessy}, {O'Sullivan}, {Ochs}, {Ogle}, {Oleszczuk},
  {Olmsted}, {Osborne}, {Ottens}, {Owens}, {Pacifici}, {Pagan}, {Page},
  {Parrish}, {Patapis}, {Pauly}, {Pavlovsky}, {Pedder}, {Peek},
  {Pena-Guerrero}, {Pennanen}, {Perez}, {Perna}, {Perriello}, {Phillips},
  {Pietraszkiewicz}, {Pinaud}, {Pirzkal}, {Pitman}, {Piwowar}, {Platais},
  {Player}, {Plesha}, {Pollizi}, {Polster}, {Pontoppidan}, {Porterfield},
  {Proffitt}, {Pueyo}, {Pulliam}, {Quirt}, {Quispe Neira}, {Ramos Alarcon},
  {Ramsay}, {Rapp}, {Rapp}, {Rauscher}, {Ravindranath}, {Rawle}, {Regan},
  {Reichard}, {Reis}, {Ressler}, {Rest}, {Reynolds}, {Rhue}, {Richon},
  {Rickman}, {Ridgaway}, {Ritchie}, {Rix}, {Robberto}, {Robinson}, {Robinson},
  {Robinson}, {Rock}, {Rodriguez}, {Rodriguez Del Pino}, {Roellig}, {Rohrbach},
  {Roman}, {Romelfanger}, {Rose}, {Roteliuk}, {Roth}, {Rothwell}, {Rowlands},
  {Roy}, {Royer}, {Royle}, {Rui}, {Rumler}, {Runnels}, {Russ}, {Rustamkulov},
  {Ryden}, {Ryer}, {Sabata}, {Sabatke}, {Sabbi}, {Samuelson}, {Sappington},
  {Sargent}, {Sauer}, {Scheithauer}, {Schlawin}, {Schlitz}, {Schmitz},
  {Schneider}, {Schreiber}, {Schulze}, {Schwab}, {Scott}, {Sembach},
  {Shaughnessy}, {Shaw}, {Shawger}, {Shay}, {Sheehan}, {Shen}, {Sherman},
  {Shiao}, {Shih}, {Shivaei}, {Sienkiewicz}, {Sing}, {Sirianni},
  {Sivaramakrishnan}, {Skipper}, {Sloan}, {Slocum}, {Slowinski}, {Smith},
  {Smith}, {Smith}, {Smith}, {Snyder}, {Soh}, {Sohn}, {Soto}, {Spencer},
  {Stallcup}, {Stansberry}, {Starr}, {Starr}, {Stewart}, {Stiavelli},
  {Straughn}, {Strickland}, {Stys}, {Summers}, {Sun}, {Sunnquist}, {Swade},
  {Swam}, {Swaters}, {Swoish}, {Taylor}, {Taylor}, {Te Plate}, {Tea}, {Teague},
  {Telfer}, {Temim}, {Thatte}, {Thompson}, {Thompson}, {Thomson}, {Tikkanen},
  {Tippet}, {Todd}, {Toolan}, {Tran}, {Trejo}, {Truong}, {Tsukamoto},
  {Tustain}, {Tyra}, {Ubeda}, {Underwood}, {Uzzo}, {Van Campen}, {Vandal},
  {Vandenbussche}, {Vila}, {Volk}, {Wahlgren}, {Waldman}, {Walker}, {Wander},
  {Warfield}, {Warner}, {Wasiak}, {Watkins}, {Weilert}, {Weiser}, {Weiss},
  {Weissman}, {Welty}, {West}, {Wheate}, {Wheatley}, {Wheeler}, {White},
  {Whiteaker}, {Whitehouse}, {Whiteleather}, {Whitman}, {Williams}, {Willmer},
  {Willoughby}, {Wilson}, {Wirth}, {Wislowski}, {Wolf}, {Wolfe}, {Wolff},
  {Workman}, {Wright}, {Wu}, {Wu}, {Wymer}, {Yates}, {Yates}, {Yeager},
  {Yerger}, {Yoon}, {Young}, {Yu}, {Zak}, {Zeidler}, {Zhou}, {Zielinski},
  {Zincke}, \& {Zonak}}]{Rigby2022}
{Rigby}, J., {Perrin}, M., {McElwain}, M., {et~al.} 2022, arXiv e-prints,
  arXiv:2207.05632.
\newblock \doarXiv{2207.05632}

\bibitem[{{Roberts-Borsani} {et~al.}(2022){Roberts-Borsani}, {Morishita},
  {Treu}, {Brammer}, {Strait}, {Wang}, {Bradac}, {Acebron}, {Bergamini},
  {Boyett}, {Calabr{\'o}}, {Castellano}, {Fontana}, {Glazebrook}, {Grillo},
  {Henry}, {Jones}, {Malkan}, {Marchesini}, {Mascia}, {Mason}, {Mercurio},
  {Merlin}, {Nanayakkara}, {Pentericci}, {Rosati}, {Santini}, {Scarlata},
  {Trenti}, {Vanzella}, {Vulcani}, \& {Willott}}]{RobertsBorsani2022}
{Roberts-Borsani}, G., {Morishita}, T., {Treu}, T., {et~al.} 2022, arXiv
  e-prints, arXiv:2207.11387.
\newblock \doarXiv{2207.11387}

\bibitem[{{Schlawin} {et~al.}(2020){Schlawin}, {Leisenring}, {Misselt},
  {Greene}, {McElwain}, {Beatty}, \& {Rieke}}]{Schlawin2020}
{Schlawin}, E., {Leisenring}, J., {Misselt}, K., {et~al.} 2020, \aj, 160, 231,
  \dodoi{10.3847/1538-3881/abb811}

\bibitem[{{Schreiber} {et~al.}(2017){Schreiber}, {Pannella}, {Leiton}, {Elbaz},
  {Wang}, {Okumura}, \& {Labb{\'e}}}]{schreiber2017}
{Schreiber}, C., {Pannella}, M., {Leiton}, R., {et~al.} 2017, \aap, 599, A134,
  \dodoi{10.1051/0004-6361/201629155}

\bibitem[{{S\'ersic}(1968)}]{Sersic1968}
{S\'ersic}, J.~L. 1968, {Atlas de galaxias australes} (Cordoba, Argentina:
  Observatorio Astronomico, 1968)

\bibitem[{{Treu} {et~al.}(2022){Treu}, {Roberts-Borsani}, {Bradac}, {Brammer},
  {Fontana}, {Henry}, {Mason}, {Morishita}, {Pentericci}, {Wang}, {Acebron},
  {Bagley}, {Bergamini}, {Belfiori}, {Bonchi}, {Boyett}, {Boutsia},
  {Calabr{\'o}}, {Caminha}, {Castellano}, {Dressler}, {Glazebrook}, {Grillo},
  {Jacobs}, {Jones}, {Kelly}, {Leethochawalit}, {Malkan}, {Marchesini},
  {Mascia}, {Mercurio}, {Merlin}, {Nanayakkara}, {Nonino}, {Paris},
  {Poggianti}, {Rosati}, {Santini}, {Scarlata}, {Shipley}, {Strait}, {Trenti},
  {Tubthong}, {Vanzella}, {Vulcani}, \& {Yang}}]{Treu2022}
{Treu}, T., {Roberts-Borsani}, G., {Bradac}, M., {et~al.} 2022, \apj, 935, 110,
  \dodoi{10.3847/1538-4357/ac8158}

\end{thebibliography}
\bibliographystyle{aasjournal}


\end{document}